\documentstyle[12pt]{article}
\renewcommand{\theequation}{\thesection.\arabic{equation}}
\newlength{\extraspace}
\newlength{\extraspaces}
\newcommand{\be}{\begin{equation}
\addtolength{\abovedisplayskip}{\extraspaces}
\addtolength{\belowdisplayskip}{\extraspaces}
\addtolength{\abovedisplayshortskip}{\extraspace}
\addtolength{\belowdisplayshortskip}{\extraspace}}
\newcommand{\ee}{\end{equation}}
 
\newcommand{\ba}{\begin{eqnarray}
\addtolength{\abovedisplayskip}{\extraspaces}
\addtolength{\belowdisplayskip}{\extraspaces}
\addtolength{\abovedisplayshortskip}{\extraspace}
\addtolength{\belowdisplayshortskip}{\extraspace}}
\newcommand{\ea}{\end{eqnarray}}
 
\newcommand{\bas}{\begin{eqnarray*}
\addtolength{\abovedisplayskip}{\extraspaces}
\addtolength{\belowdisplayskip}{\extraspaces}
\addtolength{\abovedisplayshortskip}{\extraspace}
\addtolength{\belowdisplayshortskip}{\extraspace}}
\newcommand{\eas}{\end{eqnarray*}}
 
\newcounter{subequation}[equation]
\makeatletter
\expandafter\let\expandafter\reset@font\csname reset@font\endcsname
\def\subeqnarray{\arraycolsep1pt
    \def\@eqnnum\stepcounter##1{\stepcounter{subequation}
        {\reset@font\rm(\theequation\alph{subequation})}}\eqnarray}

\makeatother
 
\newenvironment{theorem}[1]
{\vspace{3mm}\noindent {\bf #1 :} }{\vspace{2mm}}
\newcommand{\bt}[1]{\begin{theorem}{#1}}
\newcommand{\et}{\end{theorem}}
 
\newcommand{\newsection}[1]{
\vspace{12mm}
\pagebreak[3]
\addtocounter{section}{1}
\setcounter{equation}{0}
\setcounter{subsection}{0}
 
\begin{flushleft}
{\large\bf \thesection. #1}
\end{flushleft}
\nopagebreak
\medskip
\nopagebreak}

\newcommand{\NP}[1]{Nucl.\ Phys.\ {\bf #1}}
\newcommand{\PL}[1]{Phys.\ Lett.\ {\bf #1}}

\newcommand{\PR}[1]{Phys.\ Rev.\ {\bf #1}}
\newcommand{\PRL}[1]{Phys.\ Rev.\ Lett.\ {\bf #1}}

\newcommand{\intd}{\int \!\mbox{d}^4 x}
\newcommand{\is}{\! & \! = \! & \!}

\renewcommand{\d}{{\partial}}

\begin{document}
%
\begin{titlepage}
%
\renewcommand{\thefootnote}{\fnsymbol{footnote}}
\begin{flushright}
BN-TH-98-14\\
NTZ 17/98\\
hep-th/9807088
\end{flushright}
\vspace{0.8cm}
 
\begin{center}
{\Large {\bf Gauge parameter dependence in the background field gauge and 
the 
construction of an invariant charge }}
{\makebox[1cm]{  }       \\[1cm]
{\bf Rainer H\"au\ss ling$^{*}$, Elisabeth Kraus$^{\dagger}$, 
Klaus Sibold$^{*}$}\\ [3mm]
$^*${\small\sl  Institut f\"ur Theoretische Physik und} \\
{\small\sl Naturwissenschaftlich-Theoretisches 
           Zentrum, Universit\"at Leipzig} \\
{\small\sl Augustusplatz 10/11, D-04109 Leipzig, Germany} \\[0.5cm]
$^\dagger${\small\sl Physikalisches Institut, Universit\"at Bonn,} 
{\small\sl Nu\ss allee 12, D-53115 Bonn, Germany}} 
\vspace{1cm}
 
{\bf Abstract}
\end{center}
\begin{quote}
By using the enlarged BRS transformations we control the gauge  parameter 
dependence of Green functions in the background field gauge. 
We show that it is
unavoidable -- also if we consider the local Ward identity -- to introduce
the normalization gauge parameter $\xi_o$, which enters the Green
functions of higher orders similarly to the
normalization point $\kappa$.
The dependence of Green functions on $\xi_o$
is  governed 
by a further partial differential equation.
By modifying the Ward identity we are able to construct in
1-loop order a gauge parameter independent combination of 2-point vector and 
background vector functions. By  explicit construction of the next orders
we show that this combination can be used to construct a 
gauge parameter independent RG-invariant charge. However, it is seen that
this RG-invariant charge does not satisfy the differential equation of
the normalization gauge parameter $\xi_o$, and, hence, is not 
$\xi_o$-independent as required.
\end{quote}
\renewcommand{\thefootnote}{\arabic{footnote}}
\setcounter{footnote}{0}
{\begin{tabbing}
{\small PACS number(s):} \= \kill
{\small PACS number(s):} \> {\small 11.10.Gh, 11.15.-q} \\
{\small Keywords:}             \> {\small Quantum Field Theory, 
                                                          Background Field Gauge,} \\
                                          \> {\small Local Ward Identity, Invariant Charge} \\
\end{tabbing}}
\end{titlepage}
%

\newcommand{\Cal}{{\cal C}}
\renewcommand{\l}{\lambda}
\renewcommand{\a}{\alpha}
\renewcommand{\b}{\beta}
\renewcommand{\d}{\delta}
\renewcommand{\k}{\kappa}
\newcommand{\ld}{\buildrel \leftarrow \over \d}
\newcommand{\rd}{\buildrel \rightarrow \over \d}
\newcommand{\e}{\eta}
\renewcommand{\o}{\omega}
\newcommand{\dem}{\d_\o}
\newcommand{\p}{\partial}
\newcommand{\pmu}{\p_\mu}
\newcommand{\pmo}{\p^\mu}
\newcommand{\pnu}{\p_\nu}
\newcommand{\s}{\sigma}
\renewcommand{\r}{\rho}
\newcommand{\bpsi}{\bar\psi}
\newcommand{\dslash}{\p\llap{/}}
\newcommand{\pslash}{p\llap{/}}
\newcommand{\ve}{\varepsilon}
\newcommand{\uvi}{\underline{\varphi}}
\newcommand{\vi}{\varphi}
\newcommand{\ue}{u^{(1)}}
\newcommand{\Am}{A_\mu}
\newcommand{\An}{A_\nu}
\newcommand{\Fmnu}{F_{\mu\nu}}
\newcommand{\Fmno}{F^{\mu\nu}}
\newcommand{\Ga}{\Gamma}
\newcommand{\Gao}{\Gamma^{(o)}}
\newcommand{\Gae}{\Gamma^{(1)}}
\newcommand{\Gacl}{\Gamma_{cl}}
\newcommand{\Gagf}{\Gamma_{{\rm g.f.}}}
\newcommand{\Gainv}{\Gamma_{\rm inv}}
\newcommand{\pvi}{\partial\varphi}
\newcommand{\om}{{\bf w}}
\newcommand{\mn}{\mu\nu}
\newcommand{\Tmn}{T_{\mn}}
\newcommand{\Gmn}{\Ga_{{\mn}}}
\newcommand{\hT}{\hat T}
\newcommand{\emt}{energy-mo\-men\-tum ten\-sor}
\newcommand{\eit}{Ener\-gie-Im\-puls-Ten\-sor}
\newcommand{\Tc}{T^{(c)}}
\newcommand{\Tcr}{\Tc_{\rho\sigma}(y)}
\newcommand{\ha}{{1\over 2}}
\newcommand{\dalam}{{\hbox{\frame{6pt}{6pt}{0pt}}\,}}
\newcommand{\wtm}{{\bf W}^T_\mu}
\newcommand{\nnp}{\nu'\nu'}
\newcommand{\mnp}{\mu'\nu'}
\newcommand{\emn}{\eta_{\mn}}
\newcommand{\mpm}{\mu'\mu'}
\newcommand{\tbw}{\tilde{\bf w}}
\newcommand{\tw}{\tilde w}
\newcommand{\hw}{\hat{\bf w}}
\newcommand{\bw}{{\bf w}}
\newcommand{\bW}{{\bf W}}
\newcommand{\ubW}{\underline{\bW}}
\newcommand{\hmn}{h^{\mn}}
\newcommand{\gmn}{g^{\mn}}
\newcommand{\ga}{\gamma}
\newcommand{\Gf}{\Gamma_{\hbox{\hskip-2pt{\it eff}\hskip2pt}}}
\newcommand{\T}{\buildrel o \over T}
\newcommand{\Lf}{{\cal L}_{\hbox{\it eff}\hskip2pt}}
\newcommand{\np}{\not\!\p}
\newcommand{\lp}{\partial\llap{/}}
\newcommand{\ah}{{\hat a}}
\newcommand{\han}{{\hat a}^{(n)}}
\newcommand{\hak}{{\hat a}^{(k)}}
\newcommand{\dv}{{\d\over\d\vi}}
\newcommand{\zze}{\sqrt{{z_2\over z_1}}}
\newcommand{\zez}{\sqrt{{z_1\over z_2}}}
\newcommand{\Hmn}{H_{\(\mn\)}}
\newcommand{\hfrac}[2]{\hbox{${#1\over #2}$}}  
\newcommand{\smdm }{\underline m \p _{\underline m}}
 \newcommand{\tsmdm }{\underline m \tilde \p _{\underline m}} 
 \newcommand{ \Wh }{{\hat {\bf W}}^K}
\newcommand{ \CS }{Callan-Symanzik}
\newcommand{ \G}{\Gamma}
\newcommand{ \bl }{\b _ \l}
\newcommand{ \kdk }{\k \p _\k}
\newcommand{\mdm }{m \p _m} 
\newcommand{ \te }{\tau_{\scriptscriptstyle 1}}
\newcommand{ \mhi }{m_H}
\newcommand{ \mf }{m_f}
\newcommand{\bare}{^o}
\newcommand{\Pol}{{\mathbf P}}
\newcommand{\brs}{{\mathrm s}}
\newcommand{\cw}{\cos \theta_W}
\newcommand{\cws}{\cos^2 \theta_W}
\newcommand{\sw}{\sin \theta_W}
\newcommand{\sws}{\sin^2 \theta_W}
\newcommand{\cg}{\cos \theta_G}
\newcommand{\sg}{\sin \theta_G}
\newcommand{\cwg}{\cos (\theta_W- \theta_G)}
\newcommand{\swg}{\sin (\theta_W- \theta_G)}
\newcommand{\cv}{\cos \theta_V}
\newcommand{\sv}{\sin \theta_V}
\newcommand{\cvg}{\cos (\theta_V- \theta_G)}
\newcommand{\svg}{\sin (\theta_V- \theta_G)}
\newcommand{\fsc}{{e^2 \over 16 \pi ^2}}
\newcommand{\cvt}{\cos \Theta^V_3}
\newcommand{\svt}{\sin \Theta^V_3}
\newcommand{\cvf}{\cos \Theta^V_4}
\newcommand{\svf}{\sin \Theta^V_4}
\newcommand{\cgt}{\cos \Theta^g_3}
\newcommand{\sgt}{\sin \Theta^g_3}



\newsection{Introduction}

Gauge invariance is an issue of central importance 
in gauge theories: physical quantities
have to be gauge invariant. In perturbative calculations
gauge invariance is to be
accompanied by gauge parameter independence. 
A quantity qualifies as observable
only if it is gauge invariant and independent of gauge 
parameters used for the construction
of Green functions. In concrete calculations one often 
uses gauge parameter independence
as a hint that an object under study might indeed be physical 
(although gauge parameter
independence clearly is only a necessary condition). In practice 
the discussion of 
(in)dependence is inevitably linked with questions of 
scheme dependence - meaning two
different things: the way one is removing divergencies 
and the normalization conditions
one chooses for fixing the free parameters.\footnote{These 
two questions can interfere
with each other. Minimal subtractions, for instance, do both: 
remove divergencies 
{\it and} set implicitly normalization conditions.}\\
If one formulates theories purely with the help of Ward 
identities, Slavnov-Taylor identities
and the like - as we shall do below - one proceeds 
independently from any scheme and
thus has disentangled these questions. Making 
in addition vary the gauge parameters
(into Grassmann variables) \cite{KLZU75}
and taking that as a contribution to the Slavnov-Taylor
identity one can also control gauge parameter dependence 
algebraically, i.e. independently
from any scheme \cite{PISO84}.

We shall perform such a study when also background 
gauge fields are present and - in order
not to be academic - treat a well-defined problem: 
the construction of an invariant charge
in Yang-Mills theory. The use of background fields \cite{BFM, AB81}
has proven to be fruitful: once
the Slavnov-Taylor identity has been established 
(or is considered to hold in a naive form) 
one can simplify the calculation of renormalization 
constants because one is able to dispose
over a fairly naive local Ward identity (see e.g. \cite{DEDI94}). 
Equalities among
these constants suggest e.g. to translate 
properties of the effective charge from
QED to Yang-Mills \cite{WA96}. Findings in this 
direction have, of course, to be compared
with results of the pinch technique \cite{Pinch, DESI94} which yields 
gauge parameter independent
self-energies, thus candidates for observables 
and potentially having the
interpretation of an invariant charge \cite{WA96}. 
Since one is, however, manipulating diagrams
explicitly it is not obvious how these notions can be extended 
to higher orders.
Similarly in the work of Watson \cite{WA96} a specific (ghost free) 
gauge is chosen, such that there the ``analytic-algebraic'' scheme 
independent characterization of an effective
charge is missing, too.

Before describing our procedure we would 
like to point out that the existence of 
an effective charge is very important indeed. 
In renormalon calculations it is 
tacitly assumed \cite{REN}; it enters in all questions 
of improvement and scheme
dependence of perturbative contributions of finite 
order, and, in fact, becomes there
an issue of experimental relevance: if one ``improves'' 
the explicit calculation
of some finite order by replacing the coupling by the 
running coupling and compares
with experiment one faces the question of gauge 
parameter dependence. And thus a gauge
independent effective charge would obviously be of practical relevance.

The present paper is structured as follows: 
First we recall the notion of the effective
charge in QED by stressing those points which 
are needed in an extension to the
nonabelian case (section~2). 
Next we look at the local Ward identity of Yang-Mills 
theory and 
introduce external fields (background fields) in 
such a way that the correspondingly
changed Ward identity can be proved to all 
orders (sections~3,~4,~5). By varying the
gauge parameter and taking this variation into 
account in the Slavnov-Taylor 
identity we provide the basis for the construction 
 of an effective charge made out
of 2-point functions. We show that such an object  
can be constructed
(scheme independently) as being gauge independent 
and also invariant under the
renormalization group (sections~6,~7). But it turns 
out that it does not
satisfy an independence equation with respect to a 
normalization gauge parameter
$\xi_0$ (section~8) which has to be introduced in 
the course of defining the theory.
We end with some conclusions (section~9).
\newsection{Properties and construction of an invariant charge}

In renormalized perturbation theory
the QED Green functions to all  orders are uniquely determined
by the gauge Ward identity
\begin{equation}
\label{qedwi}
  \bigl( e  {\mathbf w}_{em} - \partial^\nu {\delta \over \delta A^\nu} \bigr)
  \Ga = - {1\over \xi} \Box \partial A
\end{equation}
with
\begin{equation}
\label{qedwo}
  {\mathbf w}_{em} = -i \sum_{f} Q_f \Bigl( \bar{f} 
  \frac{\stackrel{\rightarrow}{\delta}}{\delta \bar{f}} -
  \frac{\stackrel{\leftarrow}{\delta}}{\delta f} f \Bigr)
\end{equation}
and normalization conditions to be imposed on the photon and 
fermion self-energy in order to fix masses and residua of the 
propagators. Explicitly we impose:
\begin{eqnarray}
\label{qednorm1}
  \Ga^T _{AA} (p^2 = 0) = 0 & \qquad & 
  \frac 1 {p^2} \Ga^T_{AA} (p^2) \Big|_ {p^2 = \kappa^2} = 1 \\
\label{qednorm2}
  \Ga_{\bar f f}  \Big| _{\pslash = m_f} = 0 & \qquad &
  \partial _{\pslash} \Ga_{\bar f f}  \Big|_{\pslash = \kappa} = 1 
\end{eqnarray}
 $\Ga$ denotes the generating functional of 1 particle irreducible (1PI) 
Green functions. 
From
$\Ga$ the vertex functions are determined by differentiation with respect
to the classical fields, for example:
\begin{equation}
  \Ga_{A^ \mu A^ \nu} (x,y) \equiv {\delta^ 2 \Ga \over
  \delta A^ \mu(x) \delta A^\nu(y)}
\end{equation}
In (\ref{qednorm1}) we have decomposed the vector boson 2-point 
function into the
transversal part $\Ga^T_{AA}(p^2)$ and the longitudinal part 
$\Ga^L_{AA}(p^2)$:
\begin{equation}
  \Ga_{A^\mu A^\nu} = (\eta^{\mu \nu} - \frac {p^\mu p^\nu } {p^2} )
  \Ga^T_{AA} (p^2) + \frac {p^\mu p^\nu } {p^2} \Ga^L_{AA} (p^2)
\end{equation}
In the perturbative expansion
the lowest order of the generating functional of 1PI Green functions is
the classical action:
\begin{eqnarray}
  \Ga_{cl} &=& \intd \left\{ -\frac 14 F^ {\mu \nu} F_{\mu \nu}  - \frac 1{2\xi}
  (\p A) ^ 2 \right. \\
  &  & \phantom{\intd} \left. + \sum_f \bigl( i \bar f \ga^ \mu \p_\mu f 
  - m_f \bar f f + e Q_f\bar f \ga ^\mu A_\mu f \bigr) \right\}
\end{eqnarray}

For all further considerations it is important to note that
the normalization of the coupling constant $e$ is  determined by the
Ward identity. As a consequence, in invariant schemes
the wave-function normalization $z_A$
of the photon field is related to the normalization $z_e$ of the
coupling constant according to the well-known relation
\begin{equation}
\label{cwrel}
  z ^2 _e \, z_A =1 \; \; \; .
\end{equation}
In the general approach, which does not rely on properties of invariant
schemes, but just on properties of finite renormalized
Green functions, the scheme dependent relation (\ref{cwrel}) is reflected 
in relations between
the coefficient functions of the Callan-Symanzik (CS) equation,
\begin{equation}
  \Biggl( \kappa \partial_\kappa + m_f \partial_{m_f}
  + \beta_e e  \partial_e - \ga_A \biggl( \intd \Bigl(
 A {\delta \over \delta A}
  \Bigr) + 2 \xi \partial _\xi \biggr) - \ga_f \intd
  \Bigl( \bar f {\stackrel{\rightarrow}{\delta} \over \delta \bar f} +
  {\stackrel{\leftarrow}{\delta} \over \delta  f} f \Bigr) \Biggr) 
  \Ga = [\Delta_m]_3^3 \cdot \Ga \; ,
\end{equation}
and the renormalization group (RG) equation,
\begin{equation}
\label{RGqed}
  \Biggl(  \kappa \partial_{\kappa}
  + \tilde \beta_e e  \partial_e - \tilde\ga_A \biggl( 
  \intd \Bigl( A {\delta \over \delta A} \Bigr)
  + 2 \xi \partial _\xi \biggr) - \tilde \ga_f \intd
  \Bigl( \bar f {\stackrel{\rightarrow}{\delta} \over \delta \bar f} +
 {\stackrel{\leftarrow}{\delta} \over \delta  f} f \Bigr) \Biggr)
 \Ga = 0 \; \; \; ,
\end{equation}
with
\begin{equation}
  \tilde \beta_e = \beta_e \, , \quad \tilde \ga_A = \ga_A \quad \hbox{if}
  \quad
  \kappa^2 \to -\infty \quad .
\end{equation}
Indeed, by applying these equations to the Ward identity (\ref{qedwi})
one obtains that the
anomalous dimension of the photon is equal to the $\beta$-function of the
respective equation:
\begin{equation}
  \beta_e = \ga_A  \qquad \tilde \beta_e = \tilde \ga_A
\end{equation}

Finally, gauge parameter independence of the $\beta$-functions 
$\beta_e $ and $\tilde \beta_e$ and also of
the transversal photon 2-point function is derived from
the Ward identity and the normalization condition
imposed on the photon self-energy (\ref{qednorm1}).
Therefore the inverse transversal photon self-energy is an appropriate
object for defining the invariant charge of QED:
\begin{equation}
\label{qedinvch}
  Q_e (p^2, m_f^2, \kappa^2  ) = {p^2 e^2 \over \Ga^T_{AA} (p^2) }
  = {e^2 \over 1 + \Pi(p^2)}  \quad \hbox{and} \quad
\Ga^ T_{AA} (p^2 ) \equiv p^2 (1 + \Pi(p^2) )
\end{equation}
$Q_e (p^2, m_f^ 2, \kappa^ 2)$ is gauge parameter independent,
\begin{equation}
\label{gaugeind}
  \partial_\xi Q_e (p^2, m_f^2 , \kappa^2 ) = 0 \; \; \; ,
\end{equation}
well normalized due to (\ref{qednorm1}),
\begin{equation}
  Q _e(p^2, m_f^2 , \kappa^2 ) \Big|_{p^2 = \kappa^2} = e^2 \; \; \; ,
\end{equation}
and satisfies the homogeneous RG equation:
\begin{eqnarray}
\label{homoRG}
  \Bigl( \kappa \partial_\kappa 
  + \beta_e e  \partial_e \Bigr)  Q_e(p^2, m_f^2, \k^2)
  &=& 0 
\end{eqnarray}
Applying the CS operator to the invariant charge, it is seen to satisfy also
the respective CS equation but
with a soft breaking on the right hand side:
\begin{eqnarray}
  \Bigr( \kappa \partial _\kappa + m_f \partial _{m_f}
  + \beta_e e \partial_e \Bigr) Q_e(p^2, m_f^2, \k^2)
  &=& [\Delta_m]_3^3 \cdot Q_e \stackrel{p^2 \to -\infty}\longrightarrow 0 
\end{eqnarray}
with
\begin{equation}
  [\Delta_m]_3^3 \cdot Q_e \equiv Q_e {[\Delta_m]_3^3 \cdot \Ga^T_{AA}
  \over \Ga^T _{AA}} = e^2 p^2 [\Delta_m ]_3^3 \cdot G^T_{AA}
\end{equation}
$G^T _{AA}$ is the transversal part of the full photon propagator.

The RG equation (\ref{RGqed}) 
can be solved by solving first the homogeneous equation
(\ref{homoRG}) and the characteristic equation of the coupling.
The  solution of the characteristic equations,
usually called
the running coupling, can be identified with the object $Q_e (p^2,  m_f^2, \k^2)$
(\ref{qedinvch}) defined by the photon propagator,
and for this reason
$Q_e$ serves as an invariant charge of QED.

The situation is drastically changed in nonabelian gauge theories:
Due to the fact that nonabelian gauge invariance is broken non-linearly in
the course of quantization, the relation between the $\beta$-functions
and the anomalous dimensions of the vector bosons is lost.
But even worse, it is not obvious how to define an invariant charge
in terms of off-shell QCD Green functions which can be identified with
the running coupling of QCD.
Such a construction is even not possible if one wants to define
the invariant charge by a combination
of interaction vertices and 2-point functions as it is done
in the $\phi^ 4$-theory. 
The point is, that in principle it is not possible to remove the local
gauge parameter contributions to the 2- {\it and} 3-point  
functions in 1-loop order at the same time
by means of a normalization condition.
If one would do so,  physical objects like
 $\beta$-functions and  Green functions
of invariant operators would not be gauge parameter independent
in higher orders of perturbation theory.
The   technical machinery for 
controlling  the gauge parameter in a scheme-independent way 
has been provided by the
BRS-varying gauge parameter as it was introduced in \cite{PISO84}.
For fixing notations and for  convenience 
we shortly want to summarize the results
for QCD as presented in this paper, since they are the basic ingredients
for understanding the problems concerning the construction of an invariant
charge in nonabelian gauge theories.

For the purpose of the paper we restrict ourselves to SU(N) gauge
theories with massless gauge bosons and parity conservation in the
fermion sector, as it is the case, for example, in QCD. 
Invariance under nonabelian gauge transformations
\begin{eqnarray}
\label{gaugetr}
  \delta_{\omega} A_{a \mu} & = & \partial_{\mu} \omega_a + g f_{abc} A_{b \mu} \omega_c 
  \nonumber \\
  \delta_{\omega} \Psi & = & - g \omega_a T_a \Psi
\end{eqnarray}
determines the
Yang-Mills part and the matter part of the classical action:
\begin{eqnarray}
\label{clinv}
  \Gamma_{YM} 
 & = &
  - \frac{1}{4} \intd \; G_a^{\mu \nu} G_{a \mu \nu} \nonumber \\
 \Gamma_{matter}   & = &
 \intd  \left\{ i \overline{\Psi} \gamma^{\mu} {D}_{\mu} \Psi - 
  M \overline{\Psi} \Psi \right\}
\end{eqnarray}
with
\begin{eqnarray}
\label{defkov}
  G_a^{\mu \nu} & = & \partial^{\mu} A_a^{\nu} - \partial^{\nu} A_a^{\mu} +
                                    g f_{abc} A_b^{\mu} A_c^{\nu} \nonumber \\
  D_{\mu} \Psi & = & (\partial_{\mu} + g A_{a \mu} T_a ) \Psi
\end{eqnarray}
$T_a$ denote the anti-hermitean generators
of the fundamental representation, $f_{abc}$ the structure constants
of SU(N),
\begin{equation}
  [T_a,T_b] = f_{abc} T_c \; \; \; ,
\end{equation}
and $\Psi$ are N-component massive Dirac spinors.

When quantizing the theory local gauge invariance 
(\ref{gaugetr}) has to be broken due to the necessity of adding
a gauge fixing part $\Gamma_{g.f.}$ to the classical action:
\begin{equation}
\label{gf1}
  \Gamma_{g.f.} = \intd \left\{ \frac{\xi}{2} 
  B_a B_a + B_a \partial^{\mu} A_{a \mu} \right\}
\end{equation}
Here we have introduced the $(B,\xi)$-gauges, where BRS transformations
are nilpotent on all the fields. The fields
$B_a$ are  auxiliary (Lagrange multiplier) fields with dimension
2 transforming according to the adjoint representation of SU(N).
Eliminating them via their equations of motion leads back to the
usual $\xi$-gauges.
Nonabelian gauge invariance has to be replaced
by BRS symmetry introducing the Faddeev-Popov ghosts $c_a , \bar c_a$:
\begin{eqnarray}
\label{BRStrafo}
  \brs A_{a \mu} = \partial_{\mu} c_a + g f_{abc} A_{b \mu} c_c & , &
  \brs c_a = - \frac{1}{2} g f_{abc} c_b c_c \; \; \; , \nonumber \\
  \brs \bar{c}_a = B_a & , & \brs B_a = 0 \; \; \; , \\
  \brs \Psi = - g c_a T_a \Psi & , & 
  \brs \overline{\Psi} = g c_a \overline{\Psi} T_a \nonumber
\end{eqnarray} 
Following \cite{PISO84} we also transform the gauge parameter $\xi$ 
into a Grassmann variable $\chi$:
\begin{equation}
\label{brsxi}
  \brs \xi = \chi \; \; \; , \; \; \; \brs \chi = 0
\end{equation}
Enlarging the classical action by the Faddeev-Popov part, which 
depends on the ghosts and the parameter $\chi$,
\begin{equation}
\label{fp1}
  \Gamma_{\phi \pi} 
  = \intd \left\{ - \bar{c}_a \Box c_a - g f_{abc} \bar{c}_a \partial^{\mu}
  (A_{b \mu} c_c) + \hfrac 12 \chi \bar c_a B_a \right\} \; \; \; ,
\end{equation}
the classical action is invariant under the BRS transformations
(\ref{BRStrafo}) and (\ref{brsxi}):
\begin{equation}
  \brs \Ga_{cl} = 0 \qquad , \qquad \Ga_{cl} = \Ga_{YM} +
  \Ga_{matter} + \Ga_{g.f.} + \Ga_{\phi\pi}
\end{equation}
For this reason the Green functions in higher orders
have to be defined by the Slavnov-Taylor (ST) identity, instead of 
using a gauge Ward identity,
this ST identity being the functional version of classical BRS symmetry:
\begin{equation}
\label{ST}
  {\cal S} (\Ga) = 0
\end{equation}
with
\begin{equation}
\label{slavnov1}
  {\cal S} (\Gamma ) \equiv \intd \left\{
  \frac{\delta \Gamma}
  {\delta \rho_a^{\mu}} \frac{\delta \Gamma}{\delta A_{a \mu}} +
  \frac{\delta \Gamma}{\delta \sigma_a} \frac{\delta \Gamma}{\delta c_a} +
  B_a \frac{\delta \Gamma}{\delta \bar{c}_a} +
  \Gamma \frac{\stackrel{\leftarrow}{\delta}}{\delta \Psi}
  \frac{\stackrel{\rightarrow}{\delta}}{\delta \overline{Y}} \Gamma +
  \Gamma \frac{\stackrel{\leftarrow}{\delta}}{\delta Y}
  \frac{\stackrel{\rightarrow}{\delta}}{\delta \overline{\Psi}} \Gamma 
  \right\} +
  \chi\partial _\xi \Ga
\end{equation}
$\Ga$ is the generating functional of 1PI Green functions of the 
SU(N) gauge theory. 
The external fields $\rho^\mu_a , \sigma_a, Y $ and $\overline Y $ are 
coupled to the non-linear BRS transformations in
(\ref{BRStrafo}) by adding the external field part $\Ga_{ext.f.}$ to the classical action:
\begin{equation}
  \Ga_{ext.f.} = \intd \left\{ \rho^\mu_a  \brs A_{a \mu} + \sigma_a \brs c_a
 + \brs {\overline \Psi} Y +   \overline Y  \brs { \Psi} \right\}
\end{equation}
The classical action is not only BRS-invariant but
also invariant under rigid SU(N) transformations.
In symmetric theories it is possible to require
rigid SU(N)-invariance by means of a Ward identity to all orders:
\begin{equation}
\label{gwi1}
  {\cal W}_a \Gamma = 0
\end{equation}
with
\begin{equation}
\label{wop1}
  {\cal W}_a \Gamma \equiv - \intd \left\{ \sum_{\Phi}
  f_{abc} \Phi_b \frac{\delta \Gamma}{\delta \Phi_c} +
  \Gamma \frac{\stackrel{\leftarrow}{\delta}}{\delta \Psi} T_a \Psi -
  \overline{\Psi} T_a \frac{\stackrel{\rightarrow}{\delta}}{\delta \overline{\Psi}} \Gamma +
  \Gamma \frac{\stackrel{\leftarrow}{\delta}}{\delta Y} T_a Y -
  \overline{Y} T_a \frac{\stackrel{\rightarrow}{\delta}}{\delta \overline{Y}} \Gamma \right\}
\end{equation}
In (\ref{wop1}) the sum runs over all the 
fields which lie in the adjoint representation of SU(N),
\begin{equation}
\label{sum1}
  \Phi \in \{ A_{\mu},  \rho_{\mu}, c, \sigma, \bar{c}, B \} \; \; \; .
\end{equation}

In renormalized perturbation theory
the Green functions of the nonabelian gauge theory are uniquely defined
to all orders by the Ward identity
of rigid SU(N)-invariance (\ref{gwi1}), the ST identity (\ref{ST}), by
the linear gauge fixing function
\begin{equation}
  \frac{\delta \Ga}{\delta B_a} \Big|_{\chi = 0} = \xi B_a + \partial A_a 
\end{equation}
and  normalization conditions for
fixing the remaining free parameters.
As in QED we impose 
 normalization conditions 
on the self-energies of vector bosons and fermions,
\begin{eqnarray}
\label{qcdnorm}
  \Ga^T _{AA} (p^2 = 0) = 0 & \qquad &
  \frac 1 {p^2} \Ga^T_{AA} (p^2) \Big|_ {p^2 = \kappa^2} = 1 \\
\label{qcdnorm2}
  \Ga_{\bar \Psi \Psi}  \Big| _{\pslash = M} = 0 & \qquad &
  \partial _{\pslash} \Ga_{\bar \Psi \Psi}  \Big|_{\pslash = \kappa} = 
  {\mathbf 1}  \; \; \; .
\end{eqnarray}
Here  $\Ga^ T_{AA}$ denotes the transversal part of the vector 2-point 
functions:
\begin{equation}
\label{Gatrans}
  \Ga_{A^\mu_a A^\nu_b} = 
\delta_{ab}\Bigl( (\eta^{\mu \nu} - \frac {p^\mu p^\nu } {p^2} )
  \Ga^T_{AA} (p^2) + \frac {p^\mu p^\nu } {p^2} \Ga^L_{AA} (p^2) \Bigr)
\end{equation} 
Since the ST identity does not explicitly depend on the coupling, 
these conditions have to be
supplemented by a normalization condition for the coupling.
However, having introduced the BRS-varying gauge parameter, it turns
out that there remains  one invariant counterterm to the coupling,
 which has to
be independent of the gauge parameter. Gauge parameter dependence
of the coupling counterterm is completely determined by the 
$\chi$-enlarged ST identity. If one uses for QCD
an invariant scheme like dimensional regularization
 this requirement can be fulfilled by adjusting only gauge parameter
independent  counterterms
to the gauge coupling. This is
the case for example in the MS and $\overline {\hbox{MS}}$ scheme, but
not in the momentum subtraction scheme introduced in \cite{CEGO79}.
If one wants to define the coupling in a scheme-independent way at a 
specific normalization point, one has to introduce a normalization
value $\xi_o$ \cite{PISO84} of the gauge 
parameter $\xi$, at which the coupling
is normalized,
\begin{equation}
\label{normxi0}
  \Ga_{A_{a \mu} \bar \Psi \Psi} \Big|_{p_{sym}
 = \kappa \atop \xi = \xi_o} = g 
  T_a \ga^\mu \; \; \; ,
\end{equation}
or an equivalent normalization condition imposed 
on the vector boson vertex.
Here $p_{sym}$ denotes the symmetric momentum:
\begin{equation}
  p_i^2 = p^2 \qquad \quad p_i p_j = - \hfrac 13 p^2\quad \hbox{if} \quad i \neq j
\end{equation}
In fact, having a look to explicit expressions  \cite{CEGO79,DAOS96}
it is seen that
this requirement is far from being trivial.
The local part of the 3-point vector vertex 
functions, for example, depends on the gauge parameter through
the power $\xi^3$ and this dependence
cannot be removed by the normalization condition according to the
above construction. (Another example, how $\xi_0$ enters the vertex functions,
is provided by the background field gauge and presented in (\ref{1loopxi})
of this paper.)

The important point in the construction of nonabelian gauge theories,
when using a BRS-varying gauge parameter $\xi$, is that one is
able to prove gauge parameter independence of
the gauge $\beta$-functions of the CS equation and  RG equation
to all orders of perturbation theory 
independently of a specific scheme, and finally one can
proceed to prove $\xi$-independence of Green functions of invariant
operators. 
In addition to the CS equation
\begin{eqnarray}
\label{csna}
   \Biggl( \kappa \partial_\kappa + M \partial_{M}
  + \beta_g g  \partial_g 
  &\hspace{-3mm} -&\hspace{-3mm} \ga_A 
  \biggl( \intd \Bigl( A_a {\delta \over \delta A_a}   
  - B_a {\delta \over \delta B_a} - \bar c_a {\delta \over \delta \bar c_a}
  \Bigr)
  + 2 \xi \partial _\xi \biggr) \nonumber \\
  \hspace*{-33mm}  & \hspace*{-33mm} & \hspace{-33mm}
  - \ga_f \intd
  \Bigl( \bar \Psi {\stackrel{\rightarrow}{\delta} \over \delta \bar \Psi} +
  {\stackrel{\leftarrow}{\delta} \over \delta  \Psi} \Psi \Bigr)
  - \ga_c \intd \Bigl( c_a {\delta \over \delta c_a} \Bigr) \Biggr) 
  \Ga\Big| _{ext.f.= 0 \atop \chi = 0}
  = [\Delta_m]_3^3 \cdot \Ga
\end{eqnarray}
and the RG equation
\begin{eqnarray}
\label{rgna}
  \Biggl( \kappa \partial_\kappa + 
  \tilde \beta_g g  \partial_g 
  &\hspace{-3mm} -&\hspace{-3mm} \tilde
  \ga_A \biggl( \intd \Bigl( A_a {\delta \over \delta A_a}
  - B_a {\delta \over \delta B_a} - \bar c_a {\delta \over \delta \bar c_a}
  \Bigr)
  + 2 \xi \partial _\xi \biggr) \nonumber \\
  &\hspace{-3mm} -&\hspace{-3mm}
  \tilde \ga_f \intd
  \Bigl( \bar \Psi {\stackrel{\rightarrow}{\delta} \over \delta \bar \Psi} +
  {\stackrel{\leftarrow}{\delta} \over \delta  \Psi} \Psi \Bigr)
  - \tilde \ga_c \intd \Bigl(
 c_a {\delta \over \delta c_a} \Bigr) \Biggr) \Ga\Big| _{ext.f.= 0 \atop \chi = 0}
  = 0
\end{eqnarray}
there is an additional differential equation, which governs the dependence on
the normalization value $\xi_o$ of the gauge parameter
$\xi$ \cite{PISO84}:
\begin{equation}
\label{glxi0}
  (\xi_o \p _{\xi_{o}} + \beta^{\xi_0}_g g \partial_g ) \Ga = 0
\end{equation}
with 
\begin{equation}
\label{xibeta}
  \p_\xi \beta_g = 0 \quad \p_\xi \tilde \beta_g = 0 \quad \p_\xi \beta^{\xi_0}_g = 0 
\end{equation}
to all orders.
Of course, all these $\beta$-functions are scheme-dependent in
higher orders, in particular they depend on $\kappa , M $ and
$\xi _o$.

Due to the fact that there does not exist a gauge Ward identity
the anomalous dimensions for the vector boson self energies
$\ga_A$ and $\tilde{\ga}_A$  are
not related to the $\beta$-functions of the coupling. The invariant charge
is the solution of the homogeneous RG equation and of the
characteristic equations,
\begin{equation}
  \Bigl(\kappa \p _\kappa + \beta_g g \p_g \Bigr) Q_g(p^2, \kappa^ 2, \xi_0) = 0
  \quad \mbox{with} \quad  Q_g(p^2, \kappa^ 2, \xi_0) \Big|_{p^2 = \kappa^ 2} = 
  g^2 \; \; \; ,
\end{equation}
and hence it cannot be related to a combination of vector 2-point functions.
 It has been
proposed by the Pinch technique (PT) approach to nonabelian gauge theories
\cite{Pinch, DESI94} to construct effective Green functions, which satisfy a nonabelian
gauge Ward identity. Furthermore, among other requirements, it is claimed
that the PT 2-point functions are indeed gauge parameter independent 
objects and
satisfy the RG equation of the invariant charge in the asymptotic
region. Although the PT approach looks promising,
 it is nevertheless difficult for an abstract analysis because it
cannot be  simply translated into the language of usual perturbation
theory using a classical action and Feynman rules for
systematically calculating higher order corrections.

Another way to arrive at nonabelian Ward identities has been
provided by the background field gauge \cite{BFM, AB81}. It has been pointed out,
that in the Feynman gauge the Green functions of the
background fields indeed coincide with the ones of the PT \cite{DEDI94, HA94}. 
 However
the 2-point background field functions depend on the
gauge parameter due to local contributions in 1-loop order. 
In the present 
paper we show by the algebraic method using the BRS-varying gauge parameter
that the $\xi$-dependent
 contributions indeed arise from the fact that the counterterms to
the coupling cannot be adjusted arbitrarily concerning their gauge parameter 
dependence. In the explicit calculations these properties are
ensured by using the MS scheme instead of definite normalization conditions.
Hence the 2-point background field function is not an appropriate object
for defining a RG-invariant charge.

However, finally we will show that it is possible to modify 
 the background
gauge Ward identity in such a way, that a certain object constructed
with the help of the background 2-point
functions is gauge parameter independent and satisfies the
homogeneous RG equation. 
The usual background Ward identity will only be valid at the 
normalization value $\xi_0$ of the gauge parameter $\xi$ (and
at $\chi = 0$).  We explicitly construct this invariant charge
up to 3-loop order and show that this construction is unique. 
Since we had to introduce the gauge parameter normalization $\xi_o$
we also have to consider the $\xi_o$-dependence of the invariant charge.
We claim, that the invariant charge should be $\xi_o$-independent in the
same way as it is independent of the normalization point $\kappa$ by
fulfilling the homogeneous $\xi_o$-equation.
However, when we apply the $\xi_o$-equation on the RG-invariant 
combination of 2-point functions
it  turns out, that this combination does not satisfy the homogeneous
$\xi_o$-equation, and in this way the constructed object fails to be an
invariant charge  of QCD.


\newpage
\newsection{The tree approximation and the current construction of background
field gauge}

Looking for a systematic definition of the pinch technique in higher orders,
the current algebra approach to the pinch technique 
\cite{DESI94} seems us to be the most promising approach.
Furthermore, in our opinion it is
quite related to the background gauge field due to the following
reason: If one wants
to construct insertions of gauge currents into Green functions
systematically
in quantum field theory one
necessarily ends up with the background field gauge, where one understands
the background field as an external vector field coupled to the gauge current.
In this section we want to present the current construction of background field
gauge, and, at the same time, we want to fix our notations and conventions
in the tree approximation. 

Having constructed the Green functions of nonabelian gauge theories
satisfying the ST identity and the Ward identity,
we use from the action principle that gauge invariance is
broken by insertions, which are $\brs_\Ga $-invariants:
\begin{equation}
  {\mathbf w}_a \Ga =  [\partial^\mu J_{a \mu}]_4 ^4 \cdot \Ga
\end{equation}
with
\begin{equation}
  \brs_\Ga \bigl([\partial^\mu J_{a \mu}]_4 ^4 \cdot \Ga \bigr) = 0
\end{equation}
Here ${\mathbf w}_a$ is the non-integrated version of the rigid
Ward operator (\ref{wop1}),
\begin{equation}
\label{wop1oc}
  {\mathbf w}_a \Gamma \equiv -   \sum_{\Phi}
  f_{abc} \Phi_b \frac{\delta \Gamma}{\delta \Phi_c} +
  \Gamma \frac{\stackrel{\leftarrow}{\delta}}{\delta \Psi} T_a \Psi -
  \overline{\Psi} T_a 
  \frac{\stackrel{\rightarrow}{\delta}}{\delta \overline{\Psi}} \Gamma +
  \Gamma \frac{\stackrel{\leftarrow}{\delta}}{\delta Y} T_a Y -
  \overline{Y} T_a \frac{\stackrel{\rightarrow}{\delta}}{\delta \overline{Y}}
  \Gamma \; \; \; ,
\end{equation}
and $\brs_\Ga$ denotes the linearized version of the ST operator.

In the classical approximation all the breakings are local field polynomials, and 
one finds that all $\brs_{\Ga_{cl}}$-invariants are themselves
$\brs_{\Ga_{cl}}$-variations:
\begin{equation}
\label{gaugecur}
  g {\mathbf w}_a \Ga_{cl} = \brs_{\Ga_{cl}}\partial^\mu 
  \bigl( \rho_{a \mu} + D_\mu \bar c_a\bigr)  
\end{equation}
$D^\mu$ is the covariant derivative of the adjoint representation:
\begin{equation}
\label{covder}
  D^\mu \Phi_a = \partial^\mu \Phi_a - g f_{abc} \Phi_b A^\mu _c
  \quad \; \quad  \Phi \in \{ c,  \bar{c}, B \} 
\end{equation}
Explicitly we arrive at:
\begin{equation}
\label{lwi1}
  g {\mathbf w}_a \Ga_{cl} = 
  \partial^\mu {\delta \Gacl \over \delta A_a^\mu}
  + \partial^\mu \bigl( D_\mu B_a + g f_{abc} \bar{c}_b D_\mu c_c \bigr)  
  \equiv \partial^\mu {\delta \Gacl \over \delta A_a^\mu}
  + \partial ^\mu \tilde{J}_{a \mu}
\end{equation}
From (\ref{lwi1}) we read off, 
that the nonabelian gauge Ward identity is
broken by  $B$-field
and ghost-field contributions. Since the latter ones are non-linear in
the propagating fields this breaking becomes a 
non-trivial insertion in higher orders.
Therefore, in order to be able to construct insertions of conserved currents
in BRS-invariant gauge theories,   we are going to couple the
BRS-invariant current $\tilde{J}_a^{\mu}$ 
to a set of further external vector fields, 
which turn out to be  the background fields $V_{a \mu} (x)$:
\begin{equation}
\label{clacn}
  \Gamma_{cl} \longrightarrow \Gamma_{cl} + \intd \;  \tilde{J}_a^{\mu} V_{a \mu}
\end{equation}
Since the background fields  couple to a BRS-variation they are 
 transformed under BRS into 
another set of external fields $C_a^{\mu}$ with ghost number 1 \cite{GR95},
\begin{equation}
\label{BRSbgf}
  \brs V_{a \mu} = C_{a \mu} \; \; \; , \; \; \; \brs C_{a \mu} = 0 \; \; \; .
\end{equation}

By construction we end up with the usual tree action of background field
gauges:
\begin{equation}
\label{GAbgf}
  \Ga_{cl} = \Ga_{YM}+ \Ga_{matter} + \Ga^{BF}_{g.f.} + \Ga^{BF}_{ghost}
\end{equation}
with
\begin{eqnarray}
  \Ga^{BF}_{g.f.} & = & \intd \left\{ \frac \xi 2 B_a B_a + B_a ( \partial A_a - 
  \partial V_a ) - g f_{abc} B_a A_b V_c \right\} \nonumber \\
  \Ga^{BF}_{ghost} & = &
  \intd \left\{ \hfrac 12 \chi \bar c_a B_a - \bar c_a D^V D^A
  c_a + \bar c_a D^A C_{a} \right\}
\end{eqnarray}
Here $D^V$ and $D^A $ denote the covariant derivatives with
respect to vectors (\ref{covder}) and background vectors:
\begin{equation}
D_\mu^ V \Phi _a = \p_\mu \Phi_a - g f_{abc} \Phi_b V_{c\mu}
\end{equation}

The action (\ref{GAbgf})
is invariant under the enlarged Slavnov-Taylor identity,
\begin{equation}
  {\cal S}(\Ga_{cl}) = 0
\end{equation}
with
\begin{equation}
\label{slavnov2}
  {\cal S} (\Gamma ) \equiv \intd \left\{
  \frac{\delta \Gamma}
  {\delta \rho_a^{\mu}} \frac{\delta \Gamma}{\delta A_{a \mu}} +
  \frac{\delta \Gamma}{\delta \sigma_a} \frac{\delta \Gamma}{\delta c_a} +
  B_a \frac{\delta \Gamma}{\delta \bar{c}_a} +
  C^\mu_a \frac{\delta \Gamma}{\delta V^\mu_a} +
  \Gamma \frac{\stackrel{\leftarrow}{\delta}}{\delta \Psi}
  \frac{\stackrel{\rightarrow}{\delta}}{\delta \overline{Y}} \Gamma +
  \Gamma \frac{\stackrel{\leftarrow}{\delta}}{\delta Y}
  \frac{\stackrel{\rightarrow}{\delta}}{\delta \overline{\Psi}} \Gamma 
  \right\} +
  \chi\partial _\xi \Ga \; ,
\end{equation}
and satisfies the local gauge Ward identity:
\begin{equation}
\label{wardbgf}
  \biggl( g {\mathbf w}_a - 
  \partial^\nu {\delta \over \delta A_a^\nu}
  - \partial^\nu {\delta \over \delta V_a^\nu} \biggr) \Gacl  = 0
\end{equation}
${\mathbf w}_a$ is given in (\ref{wop1oc}), but the sum over
$\Phi$ now includes also the background fields $V_{a \mu}$ and their
BRS transformations $C_{a \mu}$:
\begin{equation}
  \Phi \in \{A_{\mu} , \rho_{\mu} , V_{\mu} , C_{\mu} , c, \sigma, \bar{c}, B \}
\end{equation} 

In \cite{GR95} the ST identity (\ref{slavnov2})
and the Ward identity (\ref{wardbgf}) have been considered
for $\chi = 0$ as the defining symmetries in the procedure of
algebraic renormalization of the background field gauge. According
to the fact, that we do not only want to construct background field
Green functions but also a gauge parameter independent object
out of 2-point functions
we have to modify the local Ward identity in higher orders, but
we will refer to \cite{GR95} as far as possible.

The point, where the construction of the present paper differs from
the usual background field method can be seen already from our
approach to the tree action. Considering eq.~(\ref{gaugecur}) it is
seen immediately,
that it is by no means unique how to couple the gauge current
 to the vector fields $A^\mu_a$ and 
$V_a^\mu$. In the tree approximation different
parametrizations are only trivial field redefinitions of the form
\begin{equation}
A^\mu_a \longrightarrow A^\mu_a + \bar z V^\mu_a \, .
\end{equation}
In higher
orders, however, this freedom 
 becomes an important ingredient of the construction
when we want to fix the background 2-point function in such a way that
we  arrive at a gauge parameter independent combination to all orders.


\newsection{Normalization conditions and local invariants}

Having defined the tree approximation,
the Green functions in renormalized perturbation theory are constructed
according to a renormalization scheme,  defining symmetries and normalization
conditions. 
Due to the action principle the symmetries of lowest order
are in 1-loop order at most
broken by local polynomials:
\begin{eqnarray}
  {\cal S} (\Ga ^R ) & = & \Delta_{brs} + O(\hbar^2) \\
  \Bigl( g
  {\mathbf w}_a - \p {\delta \over \delta A_a} -\p {\delta \over \delta V_a}
  \Bigl) \Ga^R & = & \partial J_a  + O(\hbar^2) \nonumber
\end{eqnarray}
$\Ga^R$ denotes the generating functional of 
subtracted finite 1PI Green functions to be calculated in
a specific renormalization scheme such as the momentum 
subtraction scheme of BPHZL or using dimensional regularization 
with MS-subtraction of poles.

The breakings of the ST identity are restricted 
according to algebraic consistency by:
\begin{equation}
  \brs_{\Gacl} \Delta_{brs} = 0
\end{equation}
In \cite{GR95} it has been shown that all possible breakings 
of the ST identity for $\chi = 0$ have to be
BRS variations, if parity is conserved. The generalization to
$\chi \neq 0$ is straightforward and proceeds along the lines of
\cite{PISO84}:
\begin{equation}
  \Delta_{brs} = - \brs_{\Ga_{cl}} \Ga_{break}
\end{equation}
$\Ga_{break}$ has the same quantum numbers as the classical action.
For this reason we are able to absorb $\Delta_{brs}$ into noninvariant
counterterms to the classical action:
\begin{equation}
\label{Gasym}
  {\cal S} (\Ga^R + \Ga_{break}) = O(\hbar^2)
\end{equation}
These counterterms are scheme-dependent and vanish immediately in 
an invariant scheme. Furthermore, it is clear that the proof outlined above
can be extended to all orders of perturbation theory by induction in the
loop expansion.

Hence, for proceeding, we will now assume that the ST identity has 
been established to all orders by an appropriate adjusting of counterterms. 
It then remains to find the free parameters of the model,
i.e.\ the invariant counterterms $\Ga_{inv}$ carrying the quantum numbers
of the action and obeying
\begin{equation}
  \brs_{\Gacl} \Ga_{inv}  = 0 \; \; \; .
\end{equation}
They have to be fixed by normalization conditions 
and/or the local Ward identity.
Determining the general classical solution $\Ga_{cl}^{gen}$ of
\begin{equation}
  {\cal S}(\Ga_{cl}^{gen}) = 0 \quad\hbox{with} \quad
  \dim^{UV}\Ga_{cl}^{gen} \leq 4
\end{equation}
allows to find all the invariant counterterms of higher orders.
We do not want to give the details of the calculation here, 
because this calculation follows
the lines of \cite{PISO84} (concerning gauge parameter dependence)
and \cite{GR95} (concerning the ST identity with background field).

The most general solution $\Ga_{cl}^ {gen}$ 
can be decomposed, just as in the tree approximation,
into the Yang-Mills part, the matter part and the external field part. We 
further split off the gauge fixing and ghost part, the latter being later on 
trivially determined from the ghost equation of motion:
\begin{equation}
  \Ga_{cl}^{gen} = \Ga^{gen}_{YM} +
  \Ga^{gen}_{matter} + \Ga^{gen}_{ext.f.} +
  \Ga^{gen}_{g.f.} + \Ga^{gen}_{\phi\pi}
\end{equation}
We find that the YM-part and the matter part depend on the vector fields
only via the combination 
\begin{equation}
\label{glA.6}
  A^o_{\mu} = z_A A^\mu + z_V V^\mu \; \; \; ,
\end{equation}
where $z_A$ and $z_V$ are arbitrary parameters:
\begin{equation}
\label{glA.7}
  \Ga^{gen}_{YM}( A^o)  = 
  - \frac{1}{4} \intd  \; G_a^{\mu \nu}  (A^o) G_{a \mu \nu} ( A^o)
\end{equation}
with
\begin{equation}
\label{glA.8}
  G_a^{\mu \nu} (A^o) \equiv  \partial^{\mu} {A}^{o\nu}_a -
  \partial^{\nu} {A}^{o\mu}_a +
  g ^o  f_{abc} {A}^{o\mu}_b {A}^{o\nu}_c
\end{equation}
Introducing the bare fermion fields
\begin{equation}
  \Psi^o = z_F \Psi  \; \; \; ,
\end{equation}
also the matter part takes its usual covariant form:
\begin{equation}
\label{glA.12}
  \Gamma^{gen}_{matter} = \intd \left\{
  i  \overline{\Psi^o} \gamma^{\mu} \partial_{\mu} \Psi^o +
  i g^o \overline{\Psi^o} \gamma^{\mu} A^o_{a \mu} 
  T_a \Psi^o - M^o
   \overline{\Psi^o} \Psi^o \right\}
\end{equation}
The external field part does not only depend on $A^o$ but shows an explicit
dependence on the background fields $V^\mu_a$ and their BRS transformations
$C^\mu_a$:
\begin{eqnarray}
\label{gaextf}
  \Ga^{gen}_{ext.f.}& = & \intd \left\{ - \frac 12 f_{abc} z_G g^o \sigma_a c_b c_c -
  \chi \partial_\xi \ln z_G \sigma_a c_a \right. \\
  & & \phantom{\intd} \left.
 + z_A^{-1} \rho^{\mu}_a \bigl( z_G \partial_\mu c_a + g^o z_G f_{abc} A_{b \mu}^o
  c_c  - z_V C_{a \mu} ) \right. \nonumber \\
  & & \phantom{\intd} \left.
  + \chi z_A^{-1} \p_\xi \ln z_A  \rho^{\mu}_a A_{a \mu}^o + 
  \chi z_V z_A^{-1} \p_\xi \ln {z_V \over z_A}  \rho^{\mu}_a V_{a \mu} 
  \right. \nonumber \\
  & & \phantom{\intd} \left.
  - g^o (\bar Y T_a \Psi +
  \bar \Psi T_a Y) c_a z_G  + \chi \p_\xi \ln z_F  (\bar Y \Psi - \bar \Psi Y )
  \right\} \nonumber
\end{eqnarray}
Having introduced a BRS-transforming gauge parameter, it is also $\chi$-dependent.
The most remarkable point in the analysis is, however, the fact that both
the parameter $g^o$, which is usually identified with the coupling, and 
the parameter $M^o$, which represents the bare mass of fermions, have
to be gauge parameter independent in order to be consistent with the
$\chi$-enlarged Slavnov-Taylor identity:
\begin{equation}
\label{xiind}
  \partial_\xi g^o = 0 \qquad \partial _\xi M^o = 0
\end{equation}
The wave function renormalizations $z_A, z_V $ and $z_F$, on the other hand,
are allowed to depend on $\xi$ arbitrarily. 

Before proceeding to the gauge fixing sector we apply the general
local abelian Ward operator, compatible with the algebra, to
the general YM, matter and external field part and find by
direct computation:
\begin{equation}
  \Bigl( g^o (z_A \bar z + z_V ) {\mathbf w}_a - \bar z \p {\delta \over \delta
  A_a } -  \p {\delta \over \delta V_a} \Bigr)
  (\Ga_{YM}^{gen} + \Ga_{matter}^{gen} + \Ga_{ext.f.}^{gen} ) \Big|_{\chi = 0} = 0
\end{equation}
$\bar z $ is a further by now undetermined parameter. 
Acting with this Ward operator on the general linear gauge fixing function
which takes for $V= 0$ the usual form,
\begin{equation}
  \Ga_{g.f}|_{V= 0} = \intd \left\{ \frac \xi 2 B_a B_a + B_a \partial A_a \right\} \; \; \; ,
\end{equation}
it turns out that the background field dependent part is uniquely
determined from invariance with respect to the local Ward operator:
\begin{equation}
\label{gagf}
  \Ga_{g.f} = \intd \left \{ \frac \xi 2 B_a B_a + B_a \partial A_a -
  \bar z B_a \partial V_a - g^o (z_A \bar z + z_V) 
  f_{abc} B_a A^\mu_b V_{c \mu} \right\}
\end{equation}
The ghost part of the action is then uniquely given by integration of
the ghost equation of motion and is by construction
invariant under the local Ward identity:
\begin{eqnarray}
\label{gafp}
  - {\delta \over \delta \bar c_a} \Ga^{gen}_{cl} & = &
  \Bigl( \partial ^\mu \delta_{ac} +
  g^o (z_A \bar z + z_V) f_{abc} V_b^\mu \Bigr) {\delta \Ga_{cl}^{gen}
  \over \delta \rho_c^\mu}
  -  \Bigl( \partial^\mu \delta_{ac}
  \bar z + g^o(z_A \bar z + z_V) f_{abc} A_b^\mu \Bigr) C_{c \mu} \nonumber \\ 
  & & + \frac 12 \chi B_a  - 
  \chi \partial_{\xi} \bar{z} \partial^{\mu} V_{a \mu} -
  \chi g^o \partial_{\xi} (z_A \bar{z} + z_V) f_{abc} A^\mu_b V_{c \mu}
\end{eqnarray}
Hence we have
\begin{equation}
\label{Gagencl}
  \Ga_{cl}^{gen} = \Ga^{gen}_{YM} + \Ga^{gen}_{matter} +
  \Ga^{gen}_{ext.f.} + \Ga_{g.f.} + \Ga^{gen}_{\phi\pi} \; \; \; ,
\end{equation}
and the five parts of the general classical action are given in (\ref{glA.7}),
(\ref{glA.12}), (\ref{gaextf}), (\ref{gagf}) and (\ref{gafp}), respectively.\\
Finally taking into account $\chi$-dependent contributions 
to the local WI, these contributions being
uniquely constructed by requiring the Ward operator to be
a $\brs_\Ga$-invariant operator, we end up with the following local WI:
\begin{equation}
  \Bigl( g^o (z_A \bar z + z_V ) {\mathbf w}_a - \bar z \p {\delta \over \delta
  A_a } -  \p {\delta \over \delta V_a} \Bigr)  \Ga_{cl}^{gen} =
  \chi  \partial^ \mu
  \biggl(\partial_\xi \bar{z} \rho_{a \mu} -
  \partial_{\xi} \ln (z_A \bar{z} + z_V) \Bigl(
  \frac{\delta \Ga_{cl}^{gen}}{\delta C_a^\mu} + \bar{z} 
  \rho_{a \mu} \Bigr) \biggr)
\end{equation}
The local WI does not impose any further restrictions on the 
$\xi$-dependence of the free parameters of the theory.

We now want to fix the free parameters by normalization conditions.
For the vector boson and fermion self-energies we take the usual
QCD normalization conditions as given in (\ref{qcdnorm}), 
(\ref{qcdnorm2}). These
normalization conditions fix $z_A , z_F$ and $M^o$. On-shell normalization
conditions have been shown to be in agreement with the ST identity
without being forced to introduce a normalization value $\xi_o$ 
of the gauge parameter $\xi$ into these normalization
conditions \cite{HAKR97}. 
Therefore it remains to fix  the parameters $\bar z, z_V$ and
$g^o$. In a first step, we eliminate one parameter from the gauge fixing
function by requiring the gauge fixing to be
\begin{equation}
\label{gaugef3}
  \left. {\delta \Ga \over \delta B_a} \right|_{\chi = 0} = \xi B_a + \partial A_a - \bar z
  \partial V_a - g f_{abc} A_b V_c \; \; \; .
\end{equation}
 $g$ is the coupling of the tree approximation.
Hence we have:
\begin{equation}
\label{parxi}
  g^o (z_A \bar z + z_V) = g \quad \hbox{and} \quad \partial_\xi
  (z_A \bar z ) = - \partial_\xi z_V
\end{equation}
 The coupling constant is  fixed by the Ward identity 
and we remain with the 2 free parameters $z_V$ and $ \bar z$.
Due to gauge parameter independence of the coupling (\ref{xiind})
gauge parameter dependence of $z_V$ and $\bar z$ is related to the
$\xi$-dependence of $z_A$. Therefore not all of these three parameters
can be fixed arbitrarily concerning their gauge parameter dependence.
$z_A$ has already been fixed by the normalization of the vector boson residuum
(\ref{qcdnorm}).
In the usual background field approach one chooses $\bar z$ to
take a definite value such as for example $\bar z = 1$ and disposes in this way on
the gauge parameter dependence of $\bar z$. The gauge parameter dependence of
$z_V $ is then restricted and the background vector boson 2-point function can
only be normalized concerning its gauge parameter independent part.
(See (\ref{wardlocxi}), (\ref{normz_Vxi}) and (\ref{1loopxi})
for the respective Ward identity, the normalization condition
 and the explicit expression.)
For this reason the 2-point function in the usual background
field gauge depends on the gauge parameter $\xi$.
 One can, however, also proceed differently (having at our disposal the additional
 parameter $\bar z$) and dispose on $z_V$ completely by a
 normalization on the background vectors (see (\ref{normz_V})).
 But then the gauge parameter dependent part of $\bar z$ is completely determined
by the ST identity, i.e. by (\ref{parxi}), and can be fixed only at $\xi = \xi_o$.
(See (\ref{wardlocgen}) and (\ref{1loopdef2}), (\ref{1loopdef3}) for the
generalized Ward identity and explicit expressions.)

In the next section we will
establish the local Ward identity in higher orders. 
There we will show that all the breakings  of the local Ward identity can
 be absorbed by adjusting the
free parameters of the invariant counterterms, which we have computed in
the general classical approximation.

%


\newsection{The local Ward identity in higher orders}

In order to prove the local WI in higher 
orders we start from the renormalized
ST-symmetric Green functions as constructed in (\ref{Gasym}):
\begin{equation}
\label{STrsym}
  {\cal S}(\Ga^R _{sym}) = 0 + O(\hbar^2)
  \quad \hbox{with} \quad \bigl( \Ga^R_{sym} \bigr)^{(\leq 1)} =
\bigl(  \Ga^R \bigr) ^{(\leq 1)}  + \Ga^{(1)}_{break} 
\end{equation}
As for the ST identity we will proceed by induction in the 
loop expansion and give
explicitly the step from the tree approximation to 1-loop order.\\
Furthermore we postulate the linear gauge fixing function of the
tree approximation to be valid for $\Ga^R_{sym}$:
\begin{equation}
\label{gaugefix}
  \left. {\delta \Ga^R_{sym} \over \delta B_a} \right|_{\chi = 0} = 
  \xi B_a + \partial A_a -
  \tilde z \partial
  V_a - g f_{abc} A_b V_c
\end{equation}
On the gauge condition
we fix the normalization of the $B_a $-field, of the gauge parameter and
of the background field $V^\mu_a$. As long as 
a Ward identity is not used $\tilde z $ is a free 
parameter of the gauge condition.
One has also to note that the field normalization $z_V$ (\ref{glA.6}), 
which gives the mixing between the propagating field $A^\mu_a$
and the background field $V^\mu_a $, is not fixed on the gauge condition
without considering a Ward identity.
According to the discussion of the last section the Green functions 
$\Ga^R_{sym}$ are not uniquely determined by the 
subtraction scheme, the Slavnov-Taylor identity
and by requiring the linear gauge fixing, but they
can be modified by the invariant counterterms, which are related to
the free parameters of the general classical solution 
(cf.~(\ref{Gagencl})).
In perturbation theory these invariant counterterms 
are most easily expressed
in terms of invariant operators acting on the 
classical action~\cite{GR95},
\begin{eqnarray}
\label{glinvc}
  \Ga ^{(\leq 1)} \! =  \!\bigl( \Ga^R_{sym} \bigr)^{(\leq 1)}
  \hspace{-3mm}
  & + \hspace{-3mm} & \delta z_A^{(1)}
  \biggl( \intd \Bigl(
  A_{a \mu} {\delta \over \delta A_{a \mu}} - 
  \rho_{a \mu} {\delta \over \delta \rho_{a \mu}} -
  B_a {\delta \over \delta B_a} - \bar c_a {\delta \over \delta \bar c_a} \Bigr)
  + 2 \xi \partial_{\xi} + 2 \chi \partial_{\chi} \biggr) \Ga_{cl} \nonumber \\
  \hspace{-3mm}  & +\hspace{-3mm} & \chi \partial_{\xi} \delta z_A^{(1)}
  \biggl( \intd \Bigl(
  \rho_{a \mu} A_a^{\mu} - \bar{c}_a {\delta \Ga_{cl} \over \delta B_a} \Bigr)
  + 2 \xi \partial_{\chi} \Ga_{cl} \biggr) \\
  \hspace{-3mm}  & +\hspace{-3mm} & \delta z_G^{(1)} 
  \intd \Bigl( c_a {\delta \over \delta c_a} -
  \sigma _a {\delta \over \delta \sigma _a} \Bigr) \Ga_{cl} -
  \chi \partial_{\xi} \delta z_G^{(1)} \intd \;  \sigma_a c_a \nonumber \\
  \hspace{-3mm}  & + \hspace{-3mm} & \delta z_V^{(1)} \intd \Bigl( V_{a \mu} 
  {\delta \Ga_{cl} \over \delta A_{a \mu}} - 
  \rho_{a \mu} C_a^{\mu} \Bigr) + \chi \partial_{\xi} \delta z_V^{(1)} \intd \;
  \rho_{a \mu} V_a^{\mu} 
  \nonumber \\
  \hspace{-3mm}  & + \hspace{-3mm}& \delta z^{(1)}_F 
  \intd \Bigl( \overline{\Psi}
  {\stackrel{\rightarrow}{\delta} \over \delta \overline{\Psi}} +
  {\stackrel{\leftarrow}{\delta} \over \delta \Psi} \Psi -
  \overline{Y} {\stackrel{\rightarrow}{\delta} \over \delta \overline{Y}} -
  {\stackrel{\leftarrow}{\delta} \over \delta Y} Y \Bigr) \Ga_{cl} \nonumber \\
  \hspace{-3mm} & + \hspace{-3mm} & 
  \chi \partial_{\xi} \delta z^{(1)}_F \intd \Bigl(
  \overline{Y} \Psi - \overline{\Psi} Y \Bigr) \nonumber \\
  \hspace{-3mm}  & + \hspace{-3mm}& \delta g^{(1)} 
  \biggl( g \partial_g - \intd \Bigl( 
  V_{a \mu} {\delta \over \delta V_{a \mu}} + 
  C_{a \mu} {\delta \over \delta C_{a \mu}} \Bigr) \biggr) \Ga_{cl} +
  \delta M^{(1)} M \partial_{M} \Ga_{cl} \; \; \; , \nonumber
\end{eqnarray}
with:
\begin{equation}
  {\cal S} (\Ga ) = {\cal S} (\Ga^R_{sym}) = 0 + O(\hbar^2)
\end{equation}
According to (\ref{xiind}) the counterterms 
proportional to $\delta g^{(1)}$ and
$\delta M^{(1)}$ have to be gauge parameter independent:
\begin{equation}
  \partial_\xi 
  \delta g^{(1)} = 0 \qquad \quad \partial_\xi \delta M^{(1)} =0
\end{equation}
The counterterms to the gauge fixing function are already constrained
in such a way that
$\Ga $ satisfies the linear gauge condition (\ref{gaugefix}):
\begin{equation}
  \left. {\delta \Ga \over \delta B_a} \right|_{\chi = 0}  = 
  \xi B_a + \partial A_a -
  (1 + \delta \tilde z ^{(1)} -  \delta z_A ^{(1)} - \delta z_V
  ^{(1)} 
  - \delta g^{(1)}  )
  \partial
  V_a - g f_{abc} A_b V_c
\end{equation} 
By now all the free parameters can be fixed by independent normalization 
conditions.

In a first step we apply the local Ward operator 
of the tree approximation to
$\Ga^R_{sym}$. According to the action principle this WI  
is broken by local contributions $\partial^\mu J^{(1)}_{a \mu}$ with
$\phi \pi$-charge $0$ and dimension bounded by $4$:
\begin{equation}
\label{lwigarsym}
  \Bigl( g {\mathbf w}_a - \partial {\delta \over \delta A_a} -
  \partial {\delta \over \delta V_a} \Bigr) \Ga^R_{sym} = \partial^\mu
  J_{a\mu}^ {(1)} + O(\hbar^ 2)
\end{equation}
Since the 1PI Green functions  $\Ga^ R_{sym} $
are assumed to satisfy the ST identity (\ref{STrsym}),
the breakings $\partial ^\mu J^ {(1)}_{a \mu}$ have also to be 
$\brs_{\Gacl}$-invariant:
\begin{equation}
  \brs_{\Ga_{cl}}  \partial J_a^{(1)} = 0
\end{equation}
It turns out that the most general expression 
for the current $J_a^\mu$ is a $\brs_{\Gacl}$-variation:
\begin{eqnarray}
  J^\mu_a & = &\brs _{\Gacl}\Bigl( u_1 (\xi) \rho^\mu_a + u_2 (\xi)
  {\delta \Gacl \over \delta C_{a \mu}} + u_3(\xi) \partial^\mu \bar c_a
  \Bigr) \nonumber \\
\label{Jmu}
  & = &  u_1 {\delta \Gacl \over \delta A_{a \mu}} + 
  u_2 {\delta \Gacl \over \delta V_{a \mu}} + 
  u_3 \partial^\mu B _a +
  \chi \partial_\xi u_1 \rho^\mu_a 
  + \chi \partial _\xi u_2 {\delta \Gacl \over \delta C_{a\mu}} +
  \chi \partial_\xi u_3  \p^ \mu \bar c_a 
\end{eqnarray}
Testing (\ref{lwigarsym}), (\ref{Jmu}) on the gauge 
condition (\ref{gaugefix}) we find that the
coefficients $ u_2 $ and $u_3$ vanish, whereas  $ u_1$ is arbitrary
and determines the parameter $\tilde z = 1 + \delta \tilde z^ {(1)}$
which appears in the gauge fixing condition (\ref{gaugefix}):
\begin{equation}
\label{Jmugauge}
  u_2 = u_ 3 = 0 \qquad \quad  \delta \tilde z ^ {(1)} = u_1
\end{equation}

Next we apply the Ward operator of the tree approximation
to the general 1PI Green functions
$\Ga$ (\ref{glinvc}). Taking into account the results (\ref{Jmu})
and (\ref{Jmugauge}) we finally end up with:
\begin{eqnarray}
  \! &\! \! &  \!  \Bigl(g {\mathbf w}_a - \partial^ \mu 
  {\delta \over \delta A^\mu_a} -
  \partial^ \mu {\delta \over \delta V^ \mu_a} \Bigr) \Ga   \\  
  \is   \Bigl(u_1^ {(1)} 
  - \delta  z_A^ {(1)} - \delta  z_V^ {(1)}  \Bigr)
  \partial^ \mu{\delta \Gacl \over \delta A^ \mu_a}  +
  \chi \partial_ \xi \Bigl(u_1 ^ {(1)} 
  - \delta  z_A^ {(1)} - \delta  z_V^ {(1)}  \Bigr)
  \partial^ \mu \rho_{a\mu} 
  \nonumber  \\
  \! & \!  \! &  \! - \delta g^ {(1)} \Bigl (g {\bf w}_a - 
  \partial ^ \mu{\delta  \over \delta V^ \mu_a} \Bigr)\Gacl + O(\hbar^ 2 )  
  \nonumber  \\
  \is \Bigl(u_1^ {(1)} 
  - \delta  z_A^ {(1)} - \delta  z_V^ {(1)} - \delta g^ {(1)} \Bigr)
  \partial^ \mu{\delta \Gacl \over \delta A^ \mu_{a } }  +
  \chi \partial_\xi \Bigl( u_1^ {(1)} 
  - \delta  z_A^ {(1)} - \delta  z_V^ {(1)}  \Bigr)
  \partial^ \mu \rho_{a\mu} + O(\hbar^ 2) \nonumber 
\end{eqnarray}
Now we are able to follow the discussion of the last section: Having fixed
the counterterm $\delta z_A^ {(1)}$ by the normalization condition
on the residuum of the transversal vector propagator (see (\ref{qcdnorm})),
\begin{equation}
\label{normz_A}
  \frac 1  {p^2} \Ga^ T_{AA} (p^2) \Big|_{p^2 = \kappa ^2} = 1 \; \; , 
\end{equation}
and fixing the counterterm $\delta z_V^ {(1)}$ by a gauge parameter
independent normalization condition
on the external vector bosons\footnote{$\Gamma^T_{AA}$ is defined in 
(\ref{Gatrans}), and analogous expressions are
valid for $\Ga^ T_{AV}$ and $\Ga^ T_{VV}$.},
\begin{equation}
\label{normz_V}
  \frac 1 {p^2} \Bigl(\Ga^ T_{V V} + 2 \Ga^ T _{VA} + \Ga^ T_{AA}
  \Bigr) \Big|_{p^2 =
  \kappa^2 } =  1 \; \; , 
\end{equation}
the local Ward identity of the tree approximation
can only be established in 1-loop order 
concerning its $\xi$-inde\-pen\-dent
part by requiring:
\begin{equation}
  u^ {(1)}_1 - \delta z_A^ {(1)} - \delta z_V^ {(1)}\Big|_{\xi = \xi_o} = 
  \delta g^ {(1)}
\end{equation}
The $\xi$-dependent part is completely determined by the $\chi$-enlarged 
Slavnov-Taylor identity.\\
In other words: Using the normalization conditions
(\ref{normz_A}) and (\ref{normz_V}), we are able to establish the 
following local WI in 1-loop order:
\begin{equation}
\label{wardlocgen}
  \Bigl(g {\bf w}_a - \bar z \partial {\delta \over \delta A_a}
  -\partial {\delta \over \delta V_a} \Bigr) \Ga =
  \chi \partial _\xi \bar z \partial \rho_a + O(\hbar^2)
\end{equation}
with 
$$\bar z \equiv  1 + u^ {(1)} - \delta z_A^ {(1)} - \delta z_V^ {(1)}
- \delta g^ {(1)} $$
and 
$$\bar z \equiv \bar z (\xi , \xi_o ) \qquad \mbox{and} \qquad
\bar z  \Big|_{\xi = \xi_o} =1 $$
At the normalization value $\xi_0$ of the gauge parameter 
$\xi$ and at  $\chi = 0$ the 1PI
Green functions $\Ga$ satisfy the Ward identity of the tree
approximation:
\begin{equation}
  \Bigl(g {\bf w}_a - \partial {\delta \over \delta A_a}
  -\partial {\delta \over \delta V_a} \Bigr) \Ga \Big|_ {\xi = \xi_o
  \atop \chi = 0} = 0
\end{equation}
In 1-loop order we find the following explicit expressions for
the vector self-energies and the parameter $\bar z$: 
\begin{eqnarray}
\label{1loopdef1}
  \Ga^{T(1)}_{AA} &= &
  \frac {g^2}{ 16 \pi^2}\left[\left(-13 \over 16  + {\xi \over 2}
  \right) N + \frac 2 3 n_f \right] \ln \frac {p^2}{\kappa^2} \\
\label{1loopdef2}
  \Ga^{T(1)}_{AA} +2 \Ga^{T(1)}_{VA}  + \Ga^{T(1)}_{VV}  & =  &
  \frac {g^2}{ 16 \pi^2}\left[{-11 \over 3  }
  N  + \frac 2 3 n_f \right] \ln \frac {p^2}{\kappa^2} \\
\label{1loopdef3}
  \delta \bar z^{(1)} & =& \frac {g^2}{ 32 \pi^2}\left[
  \frac{(\xi -1 )(\xi + 7)} 4 N - \frac{(\xi_o -1 )(\xi_o + 7)} 4 N \right]
\end{eqnarray}

In the above construction
the relevance of algebraic gauge parameter control becomes quite
striking: The arguments we have given here follow to some extent
the arguments of~\cite{GR95}
but differ in the conclusions. Having a $\xi_o$-independent
normalization condition for $\delta z_V$ 
(\ref{normz_V}) and also for $\delta z_A$ (\ref{normz_A}) the Ward identity
of the tree approximation can only be established at $\xi = \xi_o$ 
(and $\chi = 0$) in higher orders. If one wants to establish 
the Ward identity of
the tree approximation for all values of $\xi$ 
in higher orders,
i.e.~in 1-loop order
\begin{equation}
\label{wardlocxi}
  \Bigl(g {\bf w}_a - \partial {\delta \over \delta A_a}
  -\partial {\delta \over \delta V_a} \Bigr) \Ga  = O(\hbar^2) \; \; ,
\end{equation}
the normalization condition
for the background vector has to be stated at $\xi = \xi_o$
and, accordingly,  (\ref{normz_V}) has to be modified to:
\begin{equation}
\label{normz_Vxi}
  \frac 1 {p^2} \Bigl(\Ga^ T_{V V} + 2 \Ga^ T _{VA} + \Ga^ T_{AA}
  \Bigr) \Big|_{p^2 =
  \kappa^2 \atop \xi = \xi_0} =  1 
\end{equation} 
The local part of
the combination $\Ga^ T_{V V} + 2 \Ga^ T _{VA} + \Ga^ T_{AA}$
then becomes gauge parameter dependent and is determined by the
Ward identity. Explicitly, in this case, we find the following 
gauge parameter dependent expression in 1-loop order:
\begin{eqnarray}
\label{1loopxi}
  \Ga^{T(1)}_{AA} +2 \Ga^{T(1)}_{VA}  + \Ga^{T(1)}_{VV}  & =  &
  \frac {g^2}{ 16 \pi^2}
  \left[\left({-11 \over 3  } 
  N  + \frac 2 3 n_f \right) \ln \frac {p^2}{\kappa^2}\right. \\ 
  & & \phantom{\frac {g^2}{ 16 \pi^2}}\left. + 
  \frac{(\xi -1 )(\xi + 7)} 4 N - \frac{(\xi_o -1 )(\xi_o + 7)} 4 N
  \right]
\nonumber
\end{eqnarray}

In a final step the Ward identities are established by induction 
to all orders. This is achieved by repeating the discussion given
in 1-loop order. In the following sections we will
show that we are now able to construct an RG-invariant object out
of 2-point functions, which is gauge parameter independent and which,
in lowest order, is related to the combination $\G_{AA}^ T
+2\Ga _{VA}^ T + \Ga_{VV}^ T $ (\ref{1loopdef2}) as 
suggested by the Pinch technique
construction.


\newsection{Parametric differential equations}

In this section we derive the Callan-Symanzik equation 
of the nonabelian gauge theory in the generalized
background field gauge.
In addition we will derive the differential equations, which govern
the dependence on the normalization points, namely the
renormalization group equation and the $\xi_o$-equation. (In 
(\ref{csna}) - (\ref{glxi0}) we have given the respective equations in
conventional linear gauges.) Having rigorously derived these differential
equations we can systematically proceed to construct an invariant charge
out of 2-point functions. 
According to the existence of a
Ward identity
 it is seen,
that there is a certain combination of 2-point functions, whose leading
logarithms are given by the $\beta$-functions of the coupling constant.
By introducing  the BRS-varying gauge parameter and constructing Green
functions
invariant with respect to the $\chi$-enlarged ST identity, it is ensured,
  that $\beta$-functions
are gauge parameter independent and so are the leading logarithms of
these 2-point functions (\cite{AB81, KA74}).
For this reason such an object seems to be an appropriate
object for constructing
a RG-invariant charge like in QED (see section~2, 
(\ref{gaugeind})-(\ref{homoRG})). Due to the generalized background
gauge fixing
it is even possible to remove the local $\xi$-dependent parameters
in 1-loop order (cf.~(\ref{1loopdef2})). Indeed we will be able to construct an
object which satisfies the RG equation of the invariant charge up to 3-loop
order (see section~7), but we will also
find
that this object does not satisfy the appropriate differential 
$\xi_o$-equation
(see
section~8).

In order to derive the parametric differential equations
in question we first note that the 
corresponding differential operators
$\kappa \partial_{\kappa} + M \partial_M$ (for CS),
$\kappa \partial_{\kappa}$ (for RG) and 
$\xi_0 \partial_{\xi_0}$ are BRS-symmetric and rigidly symmetric
differential operators. Hence according to the quantum action principle
\begin{equation}
\label{gl5.1}
  \underline{\lambda} \partial_{\underline{\lambda}} \Gamma =
  \Delta_{\underline{\lambda}} \cdot \Gamma
  \mbox{ for } \underline{\lambda} = \{ \kappa, M \}, \kappa,
  \xi_0
\end{equation}
the r.h. sides $\Delta_{\underline{\lambda}} \cdot \Gamma$ 
have to be symmetric insertions,
\begin{equation}
\label{gl5.2}
 \brs_{\Gamma} (\Delta_{\underline{\lambda}} \cdot \Gamma ) = 0 \qquad  \quad
{\cal W}_a (\Delta_{\underline{\lambda}} \cdot \Gamma ) = 0 \; \; ,
\end{equation}
of dimension bounded by $4$ and with $\phi \pi$-charge $0$. A basis
for these insertions has already been constructed in section~5, see 
(\ref{glinvc}).

Therefore one immediately obtains for the CS equation:
\begin{eqnarray}
\label{CS}
  & & \Biggl( \kappa \partial_{\kappa} + M \partial_M +
      \beta_g \biggl( g \partial_g - \intd \Bigl(
                   V_{a \mu} {\delta \over \delta V_{a \mu}} +
                   C_{a \mu} {\delta \over \delta C_{a \mu}} \Bigr)
              \biggr) \\
  & & -\gamma_A \biggl( \intd \Bigl( A_{a \mu} {\delta \over \delta A_{a \mu}} -
                \rho_{a \mu} {\delta \over \delta \rho_{a \mu}} -
                B_a {\delta \over \delta B_a} -
                \bar{c}_a {\delta \over \delta \bar{c}_a} \Bigr) +
                2 \xi \partial_{\xi} + 2 \chi \partial_{\chi}
                \biggr) \nonumber \\
  & & -\gamma_{\bar{V}} \intd \;  V_{a \mu} {\delta \over \delta A_{a \mu}} -
       \gamma_G \intd \Bigl(
         c_a {\delta \over \delta c_a} -
         \sigma_a {\delta \over \delta \sigma_a} \Bigr) \nonumber \\
  & & -\gamma_F \intd \Bigl(
         \overline{\Psi} {\stackrel{\rightarrow}{\delta} \over
                          \delta \overline{\Psi}} +
         {\stackrel{\leftarrow}{\delta} \over \delta \Psi} \Psi -
         \overline{Y} {\stackrel{\rightarrow}{\delta} \over
                       \delta \overline{Y}} -
         {\stackrel{\leftarrow}{\delta} \over \delta Y} Y
         \Bigr) \nonumber \\
  & & + \chi \partial_{\xi} \ga_A \biggl( \intd \Bigl(
         \bar{c}_a {\delta \over \delta B_a} \Bigr) -
        2  \xi \partial_{\chi} \biggr) \Biggr) \Gamma \nonumber \\
  & = & [\Delta_m]_3^3 \cdot \Gamma -
        \gamma_{\bar{V}} \intd \; \rho_{a \mu} C_a^{\mu} \nonumber \\
  & & + \chi \partial_{\xi} \intd \Bigl(
        \gamma_A \rho_{a \mu} A_a^{\mu} +
        \gamma_{\bar{V}} \rho_{a \mu} V_a^{\mu} -
        \gamma_G \sigma_a c_a +
        \gamma_F (\overline{Y} \Psi - \overline{\Psi} Y ) \Bigr) \nonumber
\end{eqnarray}
By means of the  BRS-varying gauge parameter we find that the 
BRS-symmetric operators which are not
BRS variations have to appear with $\xi$-independent coefficient
functions (cf.~(\ref{xibeta})),
\begin{equation}
\label{gl5.4}
  \partial_{\xi} \beta_g = 0 \; \; ,
\end{equation}
whereas all the other coefficient functions are not restricted concerning
their gauge parameter dependence.
This result
holds to all orders of perturbation theory.\\
Furthermore, testing (\ref{CS}) on the gauge fixing condition
(\ref{gaugef3}) one is able to determine $\gamma_{\bar{V}}$:
\begin{equation}
\label{gamVb}
  \gamma_{\bar{V}} = \bar{z} (\beta_g - \gamma_A) +
  (2 \gamma_A \xi \partial_{\xi} - \beta_g g \partial_g ) \bar{z}
\end{equation}
In the conventional background field approach (cf.~(\ref{wardlocxi}))
the parameter $\bar{z}$ takes the
definite value $\bar{z} = 1$ to all orders. In this case
the expression for $\gamma_{\bar{V}}$ simplifies to
\begin{equation}
  \gamma_{\bar{V}} = \beta_g - \gamma_A \quad .
\end{equation}
However, in order to have a $\xi$-independent normalization condition for the
background vector 2-point functions (cf.~(\ref{normz_V})), we
keep $\bar{z}$ as an additional free parameter of the model in
the following.

The RG equation is derived in a completely analogous manner. We just
state the final result:
\begin{eqnarray}
\label{RG}
  & & \Biggl( \kappa \partial_{\kappa} +
      \tilde{\beta}_g \biggl( g \partial_g - \intd \Bigl(
                      V_{a \mu} {\delta \over \delta V_{a \mu}} +
                      C_{a \mu} {\delta \over \delta C_{a \mu}} \Bigr)
              \biggr) \\
  & & -\tilde{\gamma}_A \biggl( \intd \Bigl( A_{a \mu} 
      {\delta \over \delta A_{a \mu}} -
                \rho_{a \mu} {\delta \over \delta \rho_{a \mu}} -
                B_a {\delta \over \delta B_a} -
                \bar{c}_a {\delta \over \delta \bar{c}_a} \Bigr) +
                2 \xi \partial_{\xi} + 2 \chi \partial_{\chi}
                \biggr) \nonumber \\
  & & -\tilde{\gamma}_{\bar{V}} \intd \; V_{a \mu} 
      {\delta \over \delta A_{a \mu}} -
         \tilde{\gamma}_G \intd \Bigl(
         c_a {\delta \over \delta c_a} -
         \sigma_a {\delta \over \delta \sigma_a} \Bigr) 
         \nonumber \\
  & & -\tilde{\gamma}_F \intd \Bigl(
         \overline{\Psi} {\stackrel{\rightarrow}{\delta} \over
                          \delta \overline{\Psi}} +
         {\stackrel{\leftarrow}{\delta} \over \delta \Psi} \Psi -
         \overline{Y} {\stackrel{\rightarrow}{\delta} \over
                       \delta \overline{Y}} -
         {\stackrel{\leftarrow}{\delta} \over \delta Y} Y
         \Bigr) \nonumber \\
  & & + \chi \partial_{\xi} \tilde{\gamma}_A \biggl( \intd \Bigl(
        \bar{c}_a {\delta \over \delta B_a} \Bigr) -
        2  \xi \partial_{\chi} \biggr) 
        \Biggr) \Gamma \nonumber \\
  & = & - \tilde{\gamma}_{\bar{V}} \intd \; \rho_{a \mu} C_a^{\mu} 
        \nonumber \\
  & & + \chi \partial_{\xi} \intd \Bigl(
        \tilde{\gamma}_A \rho_{a \mu} A_a^{\mu} +
        \tilde{\gamma}_{\bar{V}} \rho_{a \mu} V_a^{\mu} -
        \tilde{\gamma}_G \sigma_a c_a +
        \tilde{\gamma}_F (\overline{Y} \Psi - \overline{\Psi} Y ) 
        \Bigr) \nonumber
\end{eqnarray}
Please note that due to the physical normalization conditions chosen no
$\beta$-function in connection with a physical mass appears in (\ref{RG}).
Again, by means of a BRS-transforming gauge parameter, one proves the
gauge parameter independence of $\tilde{\beta}_g$,
\begin{equation}
\label{gibeta}
  \partial_{\xi} \tilde{\beta}_g = 0 \; \; ,
\end{equation}
to all orders of the perturbative expansion. The test of (\ref{RG}) on the
gauge condition determines $\tilde{\gamma}_{\bar{V}}$ in 
analogy to (\ref{gamVb}) to be:
\begin{equation}
\label{gamVbk}
  \tilde{\gamma}_{\bar{V}} = \bar{z}
  (\tilde{\beta}_g - \tilde{\gamma}_A) +
  (2 \tilde{\gamma}_A \xi \partial_{\xi} -
  \tilde{\beta}_g g \partial_g ) \bar{z}
\end{equation}

Finally, we turn to the equation describing the $\xi_0$-dependence of the
theory. Expanding $\xi_0 \partial_{\xi_0} \Gamma$ like before in the basis
of BRS-symmetric operators, some of these operators are prohibited to 
appear right from the beginning because the normalization
conditions fixing the corresponding
coefficient functions do not depend on 
$\xi_0$. Having already performed these
trivial tests on some of the normalization conditions we find:
\begin{eqnarray}
\label{xi0pre}
  & & \Biggl( \xi_0 \partial_{\xi_0} +
  \beta^{\xi_0}_g \biggl( g \partial_g -
  \intd \Bigl( V_{a \mu} {\delta \over \delta V_{a \mu}} +
  C_{a \mu} {\delta \over \delta C_{a \mu}} \Bigr) 
  \biggr) - \gamma^{\xi_0}_{\bar{V}} \intd \; 
  V_{a \mu} {\delta \over \delta A_{a \mu}} \Biggr) \Gamma \nonumber \\
  & = & - \gamma^{\xi_0}_{\bar{V}} \intd \;
  \rho_{a \mu} C_a^\mu + \chi \partial_{\xi} 
  \gamma^{\xi_0}_{\bar{V}} \intd \; \rho_{a \mu} V_a^\mu
\end{eqnarray}
Testing this equation on (\ref{normz_V}), also
$\gamma^{\xi_0}_{\bar{V}}$ has to vanish:
\begin{equation}
\label{gamVbxi}
  \gamma^{\xi_0}_{\bar{V}} = \bar{z} \beta^{\xi_0}_g -
  \beta^{\xi_0}_g g \partial_g \bar{z} -
  \xi_0 \partial_{\xi_0} \bar{z} = 0
\end{equation}
The first equality in the above equation results 
from the test of (\ref{xi0pre}) on the
gauge condition (\ref{gaugef3}). Therefore the $\beta $-function of the
$\xi_o$-equation is determined by the parameter $\bar z$:
\begin{equation}
 \beta^{\xi_0}_g =
  \bigl(\bar z - g \partial_g \bar{z}\bigr)^{-1}  
  \xi_0 \partial_{\xi_0} \bar{z} 
\end{equation}
and
\begin{eqnarray}
\label{1loopbetaxio}
 \beta^{\xi_0}_g
 =   \xi_0 \partial_{\xi_0} \delta \bar{z}^{(1)} + O(\hbar ^ 2)
&= &  - 
\frac {N g^2}{4\cdot 32 \pi^2} \xi_o \partial_ {\xi_o} \Bigl(
 (\xi_o -1 )(\xi_o + 7) \Bigr) + O(\hbar ^ 2)  
\end{eqnarray}
Here we have used the explicit result of the 1-loop order
(\ref{1loopdef3}).\\
In the background field gauge and in 1-loop order
the $\beta$-function of the $\xi_o$-equation 
is  independent of the explicit form of the 1-loop Ward identity and is
given in the normalization (\ref{normz_V}) as well as in the
conventional
normalization (\ref{normz_Vxi}) by:
\begin{equation}
\label{betaxio}
 \beta^{\xi_0}_g=  - 
\frac {N g^2}{4\cdot 32 \pi^2}2 \xi_o ( \xi_o + 3) + O (\hbar^2) 
\end{equation}

Hence using (\ref{gamVbxi}) the $\xi_0$-equation reads
\begin{equation}
\label{xi0gl}
  \Biggl( \xi_0 \partial_{\xi_0} +
  \beta^{\xi_0}_g \biggl( g \partial_g -
  \intd \Bigl( V_{a \mu} {\delta \over \delta V_{a \mu}} +
  C_{a \mu} {\delta \over \delta C_{a \mu}} \Bigr) \biggr)
  \Biggl) \Gamma = 0 \; \; ,
\end{equation}
and, again, the $\beta$-function $\beta^{\xi_0}_g$ is
$\xi$-independent to all orders, 
\begin{equation}
  \partial_{\xi} \beta^{\xi_0}_g = 0 \/,
\end{equation}
 but depends -- as seen from the explicit
expression (\ref{1loopbetaxio}) -- on $\xi_o$.

Determining the Green functions in the MS or $\overline{\mbox{MS}}$ scheme a 
gauge normalization point is not explicitly introduced, but, of course, one
has also implicitly chosen a gauge parameter value $\xi_o$  where the
background field 2-point function is normalized.
 The $\xi_o$-equation cannot be derived in
these schemes, since $\xi_o$-dependence is hidden, but nevertheless, if
we would do so, then its $\beta $-function coincides  with the one, we have
calculated in (\ref{betaxio}),  at a special value of $\xi_o$.

\newpage
\newsection{Construction of a RG equation invariant charge}

In the following we will restrict ourselves for reasons of brevity to
the massless theory in which case the RG 
equation and the CS equation coincide. The
generalization to the massive theory is easy and straightforward.\\
According to the observations (for QED) in section~2 the object
 $Q_g (p^2, \kappa^ 2,\xi_o)$ we are
looking for has (at least) to satisfy the homogeneous RG equation,
\begin{equation}
\label{RGhom}
  \Bigl( \kappa \partial_{\kappa} +
  \tilde{\beta}_g g \partial_g \Bigr) Q_g (p^2, \kappa^2, \xi_o ) = 0 \; \; ,
\end{equation}
has to be gauge parameter independent,
\begin{equation}
\label{gaugein}
  \partial_{\xi} Q_g (p^2, \kappa^2, \xi_o)= 0 \; \; ,
\end{equation}
and also well-normalized:
\begin{equation}
\label{Qnorm}
  Q _g(p^2, \kappa^2, \xi_o)
 |_{p^2 = \kappa^2} = g^2
\end{equation}
In the nonabelian theory considered in this paper, the invariant
charge $Q_g$ in addition has to obey the homogeneous $\xi_0$-equation.
 This last
requirement will be discussed in detail in the following section.\\
In a first step we will now show that an object satisfying
(\ref{RGhom})-(\ref{Qnorm}), i.e.\ a RG equation invariant charge,
can be uniquely defined in a perturbative way
out of a certain combination  of vector and background vector
self-energies.
This will be done by explicitly constructing 
the lowest orders of the self-energies $\tilde{\Pi} (p^2, \kappa^2, \xi_o)$.
 From this expression the
invariant charge is constructed as in QED, 
\begin{equation}
\label{QPi}
  Q_g(p^2 , \kappa^ 2, \xi_o) = 
\displaystyle\frac{ g^2}{1+ \tilde{\Pi}(p^2, \kappa^ 2,\xi_o)
 } \; \; , \quad
\tilde \Pi = O(\hbar) \/ \; \; ,
\end{equation}
and $\tilde \Pi$ has to satisfy the following equations
(see also (\ref{qedinvch}), (\ref{RGhom})-(\ref{Qnorm})):
\begin{eqnarray}
\label{RGhomq}
  \Bigl( \kappa \partial_{\kappa} + 
         \tilde{\beta}_g g \partial_g
         - 2 \tilde{\beta}_g \Bigr)(1+ \tilde{\Pi}(p^2, \kappa^ 2,\xi_o))
 & = & 0 \; \; , \\
\label{gaugeinq}
  \partial_{\xi} \tilde{\Pi}(p^2, \kappa^ 2,\xi_o) & = & 0 \; \; , \\
\label{qnorm}
  (1 + \tilde{\Pi} (p^2, \kappa^2, \xi_o))
  |_{p^2 = \kappa^2} & = & 1
\end{eqnarray}
Before starting with the explicit construction 
of $\tilde{\Pi}$ we want to collect
some formulae useful for the subsequent calculations
and discussions. Differentiation
of the RG equation twice with respect to $A_{a \mu}$, evaluation of
the resulting expression for all fields as well as $\chi$ set equal
to zero and projection onto the transversal part yields:
\begin{eqnarray}
\label{RGAA}
  & \kappa \partial_{\kappa} \Gamma^T_{AA} +
  \tilde{\beta}_g g \partial_g \Gamma^T_{AA} -
  2 \tilde{\gamma}_A \Gamma^T_{AA} -
  2 \tilde{\gamma}_A \xi \partial_{\xi} \Gamma^T_{AA} = 0 & 
\end{eqnarray}
( $\Gamma^T_{AA}$ is defined in (\ref{Gatrans}).)
Similarly one obtains
\begin{eqnarray}
\label{RGAV}
  & \kappa \partial_{\kappa} \Gamma^T_{AV} +
  \tilde{\beta}_g g \partial_g \Gamma^T_{AV} -
  \tilde{\beta}_g \Gamma^T_{AV} -
  \tilde{\gamma}_A \Gamma^T_{AV} -
  2 \tilde{\gamma}_A \xi \partial_{\xi} \Gamma^T_{AV} -
  \tilde{\gamma}_{\bar{V}} \Gamma^T_{AA} = 0 & \\
\label{RGVV}
  & \kappa \partial_{\kappa} \Gamma^T_{VV} +
  \tilde{\beta}_g g \partial_g \Gamma^T_{VV} -
  2 \tilde{\beta}_g \Gamma^T_{VV} -
  2 \tilde{\gamma}_A \xi \partial_{\xi} \Gamma^T_{VV} -
  2 \tilde{\gamma}_{\bar{V}} \Gamma^T_{AV} = 0 & 
\end{eqnarray}
by testing the RG equation with respect to $A_{a \mu}$, $V_{b \nu}$ or
twice with respect to $V_{a \mu}$, respectively.\\
Using (\ref{RGAA})-(\ref{RGVV}) an easy calculation shows that the
quantity $\overline{\Gamma}$, defined by
\begin{equation}
\label{gambar}
  \overline{\Gamma} = \bar{z}^2 \Gamma^T_{AA} 
  + 2 \bar{z} \Gamma^T_{AV}
  + \Gamma^T_{VV} \; \; ,
\end{equation}
obeys to all orders of perturbation theory the following homogeneous 
RG equation:
\begin{equation}
\label{gambargl}
  \kappa \partial_{\kappa} \overline{\Gamma} +
  \tilde{\beta}_g g \partial_g \overline{\Gamma} -
  2 \tilde{\beta}_g \overline{\Gamma} -
  2 \tilde{\gamma}_A \xi \partial_{\xi} \overline{\Gamma} = 0
\end{equation}
This expression is valid independently from the choice of the
parameter $\bar z$. From (\ref{1loopdef2}), (\ref{1loopdef3})
  and (\ref{1loopxi}),
respectively,
we read off that $\bar \Ga $ is gauge parameter dependent in 1-loop order
and, hence,
is
not an appropriate object for defining a RG-invariant charge.\\
We would like to mention already now that, on the other hand, 
$\overline{\Gamma}$, as defined in (\ref{gambar}), does indeed satisfy
the correct $\xi_0$-equation of an invariant charge
for arbitrary $\bar z $ (see section~8):
\begin{equation}
\label{xi0gambar}
  \xi_0 \partial_{\xi_0} \overline{\Gamma} +
  \beta_g^{\xi_0} g \partial_g \overline{\Gamma} -
  2 \beta_g^{\xi_0} \overline{\Gamma} = 0
\end{equation}

We now turn to the explicit
determination of a RG invariant charge 
 by constructing order to order in perturbation theory $\tilde \Pi(p^2,
 \kappa^ 2,\xi_o)$.
 In 1-loop order the RG equation for $\tilde \Pi$ (\ref{RGhomq})
reads:
\begin{equation}
\label{RGhomq1}
  \kappa \partial_{\kappa} \tilde{\Pi}^{(1)} 
 -    2 \tilde{\beta}_g^{(1)}  = 0
\end{equation}
A closer look to (\ref{RGAA})-(\ref{RGVV}), these equations also
written in strict 1-loop order, shows that up to local counterterms
such an object is uniquely determined by:
\begin{equation}
\label{invq1}
  p^2 \tilde{\Pi}^{(1)} = \Gamma_{AA}^{T (1)} + 2 \Gamma_{AV}^{T (1)} +
                      \Gamma_{VV}^{T (1)}
\end{equation}
This quantity is gauge parameter independent, only if we  state
the  normalization condition (\ref{normz_V}) of the generalized
background
gauge (cf.~(\ref{1loopdef2})): 
The $\kappa$-dependent part of (or the nonlocal
contributions to) $\tilde{\Pi}^{(1)}$ is $\xi$-independent since the
$\beta$-function is $\xi$-independent  (\ref{gibeta}).
Furthermore, the $\kappa$-independent part of (or the local
contributions to) $\tilde{\Pi}^{(1)}$ is also gauge parameter
independent due to the normalization condition (\ref{normz_V}).
Hence, $ 1 +
\tilde{\Pi}^{(1)}$ can be used to construct a RG-invariant charge up to
1-loop order according to eq.~(\ref{QPi}).

We want to conclude the treatment of the 1-loop approximation by
establishing the relation between $\tilde{\Pi}^{(1)}$ and
$\overline{\Gamma}^{(1)}$. A simple calculation shows that:
\begin{eqnarray}
\label{cpi1ga1}
  & p^2 \tilde{\Pi}^{(1)} = \overline{\Gamma}^{(1)} -
    2 p^2 \delta \bar{z}^{(1)} & \\
  & (\mbox{with } \bar{z} = 1 +
    \delta \bar{z}^{(1)} + \dots ) & \nonumber
\end{eqnarray}
This means that the unwanted but unavoidable gauge parameter dependence
of $\overline{\Gamma}^{(1)}$ can be (and, in fact, {\it is}) removed
by means of the local contribution $- 2 p^2 \delta \bar{z}^{(1)}$,
i.e. by means of the additional free parameter $\bar{z}$. The full power
of having at hand this additional freedom when looking for a RG equation
invariant charge, however, only turns out in higher orders.

The construction of these higher orders of $\tilde{\Pi}$ is done in
a recursive manner and follows exactly the strategy which was outlined
above for the 1-loop order. Here we skip the details of the calculation
and just present the final result for 2- and 3-loop order:
\begin{eqnarray}
\label{invq2}
  p^2 \tilde{\Pi}^{(2)} & = & \Gamma_{AA}^{T (2)} + 2 \Gamma_{AV}^{T (2)} +
                          \Gamma_{VV}^{T (2)} \\
  & & - p^2 \tilde{\Pi}^{(1)} (1 - g \partial_g ) \delta \bar{z}^{(1)} 
      + \Gamma_{AA}^{T (1)} (1 - 2 \xi \partial_{\xi} ) \delta \bar{z}^{(1)}
      \nonumber
\end{eqnarray}
\begin{eqnarray}
\label{invq3}
  p^2 \tilde{\Pi}^{(3)} & = & \Gamma_{AA}^{T (3)} + 2 \Gamma_{AV}^{T (3)} +
                          \Gamma_{VV}^{T (3)} \\
  & & - p^2 \tilde{\Pi}^{(1)} (1 - g \partial_g ) \delta \bar{z}^{(2)}
      + \Gamma_{AA}^{T (1)} (1 - 2 \xi \partial_{\xi} ) \delta \bar{z}^{(2)}
      \nonumber \\
  & & - p^2 \tilde{\Pi}^{(2)} (1 - g \partial_g ) \delta \bar{z}^{(1)}
      + \Gamma_{AA}^{T (2)} (1 - 2 \xi \partial_{\xi} ) \delta \bar{z}^{(1)}
      \nonumber \\
  & & - \frac{1}{2} \frac{1}{p^2} 
      (p^2 \tilde{\Pi}^{(1)} - \Gamma_{AA}^{T (1)}) \Bigl(
      p^2 \tilde{\Pi}^{(1)} (1 - g \partial_g ) \delta \bar{z}^{(1)} -
      \Gamma_{AA}^{T (1)} (1 - 2 \xi \partial_{\xi} ) \delta \bar{z}^{(1)}
      \Bigr) \nonumber \\
  & & - \frac{1}{2} \Gamma_{AA}^{T (1)} \tilde{\Pi}^{(1)}
      (1 + 2 \xi \partial_{\xi} )(1 - g \partial_g ) \delta \bar{z}^{(1)}
      \nonumber \\
  & & + \frac{1}{4} p^2 \tilde{\Pi}^{(1)^2}
      (1 + g \partial_g )(1 - g \partial_g ) \delta \bar{z}^{(1)} 
      + \frac{1}{4} \frac{1}{p^2} \Gamma_{AA}^{T (1)^2}
      (1 + 2 \xi \partial_{\xi} ) (1 - 2 \xi \partial_{\xi} )
      \delta \bar{z}^{(1)} \nonumber
\end{eqnarray}
The invariant charge constructed from $\tilde \Pi = 1 + \sum_{n =1}^3
\tilde \Pi^{(n)} $ is well normalized according to the normalization
conditions
(\ref{normz_A}) and (\ref{normz_V}). Gauge parameter independence is
proven order by order by using gauge parameter independence of
$\beta$-functions and the normalization conditions.
Hence, we have succeeded in defining a RG-invariant charge for
QCD (i.e. an object satisfying (\ref{RGhom})-(\ref{Qnorm})) up to
3-loop order as a unique combination of 2-point functions.


\newsection{Compatibility with the $\xi_0$-equation?}

One of the main results of the abstract construction is the observation,
that a proper scheme-independent definition 
of QCD-parameters asks for the introduction of a 
normalization gauge parameter $\xi_o$. For nonabelian gauge theories in linear
gauges the $\xi_o$ and its corresponding differential equation 
have been introduced
in \cite{PISO84} (see also section~2, (\ref{normxi0})).
In the present paper we have shown, that 
also in  the background field gauge it is not possible to avoid  the
introduction of $\xi_o$
by means of the
Ward identity
(see sections~4,~5 and 
(\ref{normz_V}), (\ref{normz_Vxi}), respectively).\footnote{We want
to mention, that  in the Abelian Higgs model
the $\chi$-enlarged
ST identity could be fulfilled without 
introducing $\xi_o$
by using the local Ward identity and on-shell conditions, see \cite{HAKR97, HAKA97}.}
The dependence of Green functions on $\xi_o $ is governed by a partial
differential equation, which ensures $\xi_o$-independence of 
physical quantities, in a similar manner as the RG equation ensures
normalization parameter independence of physical quantities.
In the background field gauge the $\xi_o$-equation was derived in 
section~6, see (\ref{xi0gl}).

It is obvious that the invariant charge should not only be
a normalization point independent object but also a $\xi_o$-independent
object, and therefore should satisfy both equations, the RG equation and the
$\xi_o$-equation, i.e.: 
\begin{equation}
\label{xi0hom}
  \left( \xi_0 \partial_{\xi_0} +
  \beta_g^{\xi_0} g \partial_g \right) Q_g (p^2, \kappa^ 2,\xi_o) = 0
\end{equation}
In terms of $1 + \tilde{\Pi}$, the ``inverse'' of $Q_g$, (\ref{xi0hom}) reads:
\begin{equation}
\label{xi0homq}
  \left( \xi_0 \partial_{\xi_0} +
  \beta_g^{\xi_0} g \partial_g -
  2 \beta_g^{\xi_0} \right) (1+ \tilde{\Pi}(p^2, \kappa^ 2,\xi_o)) = 0
\end{equation}
In the preceding section we have uniquely constructed
a gauge parameter independent RG-invariant charge in terms of $\tilde \Pi$.
An easy calculation, however, shows that this invariant charge
 fails to
fulfill (\ref{xi0homq}). In fact, one finds
\begin{equation}
\label{xi0pi1}
  \xi_0 \partial_{\xi_0} \tilde{\Pi}^{(1)} = 0  \qquad \hbox{and}
\qquad  \xi_0 \partial_{\xi_0} \tilde{\Pi}^{(2)} = 0 \quad , 
\end{equation}
i.e. the lowest orders of the RG-invariant charge does not
explicitly
depend on $\xi_o$. Since, however, the $\beta$-function 
$\beta^{\xi_0}_g$ is non-vanishing
outside
the Landau gauge, see (\ref{1loopbetaxio}), the invariant charge depends 
on $\xi_0$ via
the coupling. 

From the construction it is obvious that it is not possible to construct
a RG-invariant charge, which has well-defined normalization
conditions
and fulfills at the same time the differential $\xi_o$-equation,
 out of a combination of 2-point functions
in linear background gauges. Only the Landau gauge seems to be
distinguished
from other gauges since in this gauge the RG-$\beta$-function of the gauge
parameter $\tilde \beta_\xi = \xi \tilde \ga_A $ 
and the gauge coupling $\beta$-function
$\beta_g^{\xi_o}$ of
the $\xi_o$-equation vanish.

We want to mention that we did indeed succeed in a conventional way to
define a RG-invariant charge in 1-loop order, which is related to the
Pinch technique self-energies and  is gauge parameter independent. 
 Furthermore, we were also able
to continue this invariant charge to higher orders
(explicitly up to 3-loop). Thereby we have used  arguments similar to those
given  in the construction of Pinch technique self-energies 
in higher orders  \cite{PAPI96}.
 However, by a careful analysis of invariant
counterterms, normalization conditions and the $\chi$-enlarged
Slavnov-Taylor identity we found, that QCD has to be complemented by a
gauge parameter normalization
 and a respective partial differential equation. It
is
this equation which makes the construction of an invariant charge
impossible in the conventional linear background gauge. 
 The $\xi_o$-equation has been not considered up to now
in explicit calculations since in such calculations one frequently sticks to special
invariant
schemes like the MS and $\overline{\hbox{MS}}$ scheme. Therefore 
we claim that a first step towards a
deeper and better understanding of the construction  
of an invariant charge requires a
careful abstract solution of the RG
{\it and} the $\xi_o$-equation.
\newsection{Conclusions}

In the present paper we studied gauge parameter dependence 
in Yang-Mills theories including background fields. These were
not introduced ad hoc but rather as a tool to couple
systematically to currents which otherwise would have prohibited
the existence of a local Ward identity. As a concrete application
we tried to construct a gauge invariant charge made up from
2-point functions of the quantum and the background vector fields,
thus combining suggestions from the pinch technique with the
background field method. Up to and including three loops
we showed that such an object exists and is characterized
as:
\begin{itemize}
\item
  being solution of the homogeneous renormalization group
  equation (\ref{RGhom}),
\item
  being gauge parameter independent (\ref{gaugein}),
\item
  coinciding with the coupling at a normalization point
  (\ref{Qnorm}).
\end{itemize}
In our abstract renormalization scheme independent approach
it is, however, apparent that a complete set of normalization conditions
which is compatible with the gauge parameter dependence (controlled
algebraically to all orders) requires a normalization gauge parameter
$\xi_0$ (see sections~4,~5) which is on the same footing as the
momentum scale normalization parameter $\kappa$. It gives rise to
an analogous parametric differential equation (\ref{xi0gl}). Since
the otherwise perfect candidate for an effective charge fails
to satisfy the corresponding homogeneous equation (\ref{xi0hom})
we consider the issue of constructing an effective charge in 
Yang-Mills theories as not yet finally settled.
\newpage
{

}
%
\end{document}